\documentclass[fleqn,usenatbib]{mnras}

\usepackage{newtxtext,newtxmath}

\usepackage[T1]{fontenc}

\DeclareRobustCommand{\VAN}[3]{#2}
\let\VANthebibliography\thebibliography
\def\thebibliography{\DeclareRobustCommand{\VAN}[3]{##3}\VANthebibliography}

\usepackage{graphicx}
\usepackage{amsmath}
\usepackage{multicol}
\usepackage{bm}
\usepackage{pdflscape}
\usepackage{amsfonts}
\usepackage[nopatch]{microtype}
\usepackage{booktabs}
\usepackage[table]{xcolor}
\usepackage{multirow}
\usepackage{soul}
\usepackage{tablefootnote}

\defcitealias{NFW1995}{NFW 1995}
\defcitealias{NFW1996}{NFW 1996}
\defcitealias{NFW1997}{NFW 1997}
\defcitealias{Sullivan2024a}{S24a}
\defcitealias{Sullivan2024b}{S24b}

\title[Jet outbursts, NTP and the AGN jet duty cycle]{Jet outbursts, non-thermal pressure and the AGN jet duty cycle}

\author[A. Sullivan et al.]{
Andrew Sullivan,$^{1, 2}$\thanks{E-mail: andrew.sullivan@icrar.org}
Ross J. Turner,$^{3}$
Stanislav S. Shabala,$^{3}$
Chris Power$^{1, 2}$
and Sophie A. Young$^{3}$
\\
$^{1}$International Centre for Radio Astronomy Research, The University of Western Australia, 35 Stirling Highway, Crawley, Western Australia, 6009, Australia\\
$^{2}$ARC Centre of Excellence for All Sky Astrophysics in 3 Dimensions (ASTRO 3D)\\
$^{3}$School of Natural Sciences, University of Tasmania, Hobart, Tasmania, Australia
}

\date{Accepted XXX. Received YYY; in original form ZZZ}

\pubyear{\the\year{}}

\begin{document}
\label{firstpage}
\pagerange{\pageref{firstpage}--\pageref{lastpage}}
\maketitle

\begin{abstract}
We predict the non-thermal pressure (NTP) induced in the cores of galaxy clusters by kinetic jet feedback from an active galactic nucleus (AGN). We model a population of Fanaroff–Riley type I jets when sampling power-law distributions in jet power and age, which we evolve in time with a two-phase jet-lobe model. We couple the energy of each jet outburst to the surrounding gas inside spherical shells, allowing us to estimate the fraction of NTP to total pressure induced in the cluster. We predict the mean profile for this NTP fraction over the source population in a variety of cluster environments and for different AGN jet duty cycles. For typical gas and dark matter profiles, the mean NTP fraction peaks at $\sim 4 - 6\%$ when the AGN jets are active for $10-30\%$ of the total AGN lifecycle. These predictions are in good agreement with observational constraints, suggesting that AGN feedback imparts only small non-thermal contributions to the cluster's core. Furthermore, we find a relationship between the peak in the mean NTP fraction and the AGN jet duty cycle in a given cluster environment. Applying this to \textit{Hitomi} measurements of the NTP in the Perseus cluster, we infer an AGN jet duty cycle that is consistent with independent evidence of Perseus' AGN jet activity. We propose this as a novel approach for observationally inferring the past AGN activity of real clusters from their observed NTP fraction and environmental profiles.
\end{abstract}




\section{INTRODUCTION}

The central cores of galaxy clusters are important astrophysical environments for understanding the role of feedback from the active galactic nucleus (AGN). Two types of AGN feedback are relevant to the evolution of galaxy and cluster environments. Radiative feedback, generated by radiation emitted by matter falling onto the black-hole, will heat the surrounding gas and drive powerful winds. Kinetic feedback, generated by the relativistic streams of plasma known as `jets' that are launched from the AGN, will reach further out into the cluster and influence its larger-scale environment, generating turbulence, acceleration and uplifting of gas. This latter type of AGN feedback is expected to regulate radiative gas cooling in the cluster core, by heating and moving the gas and suppressing star formation; this heating-cooling balance is known as `maintenance-mode' feedback \citep[for an overview of AGN feedback, see][]{Fabian2012}.

Evidence for kinetic AGN feedback is seen in X-ray observations as cavities or `bubbles' in the intracluster gas, which are powered by ongoing AGN jet activity. This has been particularly well observed in the core of the Perseus cluster \citep[see, e.g.][]{Boehringer1993, Fabian2000, Churazov2000, Fabian2006}, which is the brightest X-ray galaxy cluster in the sky. The gas motion induced during these powerful outbursts will manifest as kinetic or `non-thermal' pressure (NTP hereafter). An important question concerns quantifying how significant and long-lasting this NTP is to the local pressure balance in the cluster's core.

Observational constraints on the gas kinematics, and hence the NTP, in the cores of real clusters are limited. The \textit{Hitomi} satellite directly resolved the turbulent gas velocity within the central 100~kpc of the Perseus cluster, measuring the ratio of NTP to thermal pressure as $\simeq 2-7\%$ and possibly up to $\simeq 11-13\%$ \citep{Hitomi2018}. More recently, \citet{Dupourque2023} inferred NTP from X-ray surface brightness fluctuations for a sample of galaxy clusters, estimating a NTP fraction of $\simeq 1-2\%$ in their cores. The observational consensus so far is that the lasting non-thermal contribution from kinetic AGN feedback to the cores of galaxy clusters is minimal but non-zero; however, more observations are needed. 

This general picture has been corroborated by numerical studies. \citet{BourneSijacki2017} reproduced velocity dispersions consistent with the \textit{Hitomi} results when combining jet feedback with large-scale turbulence in the intracluster gas. Moreover, \citet{Truong2024} considered simulated Perseus-like clusters subject to jet feedback (taken from the TNG-Cluster suite), which were shown to attain gas turbulence consistent with the \textit{Hitomi} results, albeit at slightly elevated levels. Modelling of the interaction between jets and their surrounding environments has improved significantly in recent years \citep[for a comprehensive overview, see][]{BourneYang2023}, but remains an ongoing field of research. Accurate AGN feedback models remain essential for informing larger cosmological simulations, which are poorly resolved at the required scales. 

NTP in galaxy clusters, more generally, will be produced by any \textcolor{black}{event that imparts a non-random velocity (e.g. turbulence, bulk flows)} to the gas. \textcolor{black}{Beyond AGN feedback, NTP can be produced by magnetic fields and cosmic rays \citep[e.g.][]{Lagana2010}, and by gravitational processes, such as shocks and mergers \citep[e.g.][]{Nelson2014}.} Hydrodynamic simulations that capture the gas turbulence induced by these gravitational processes have shown that the contribution of NTP to the overall gas pressure increases with distance from the cluster centre \citep{Nelson2014, Angelinelli2020}. Currently, observational constraints for the NTP in the cluster outskirts show significant discrepancy with these numerical predictions \citep{Siegel2018, Eckert2019, Sayers2021, Dupourque2023}. In our previous work \citep[][hereafter \citetalias{Sullivan2024b}]{Sullivan2024b}, we employed mathematical constraints on the gas entropy profile to capture the effects of gravitational processes occurring within the cluster, predicting the radial NTP profile in the cluster's outskirts. Our results from this work strongly favoured the predictions from numerical simulations.

Improving our understanding of NTP on large scales is motivated by the need to quantify its contribution to the pressure balance within galaxy clusters. Hydrostatic equilibrium is the balance between the pressure generated by the cluster's gravitational potential and the thermal pressure of its gas component; wherever NTP is present in the cluster, this pressure will also contribute to the local hydrostatic equilibrium balance. Importantly, the assumption of hydrostatic equilibrium is required to relate a cluster's X-ray emission observables to its underlying halo mass. Due to poor constraints on its profile, NTP is invariably neglected in these hydrostatic mass estimates \citep[e.g.][]{Vikhlinin2006, Vikhlinin2009}, and is expected to be the dominant contribution to the `hydrostatic bias' in the resulting mass estimates. Significant effort has been dedicated to quantifying this hydrostatic bias \citep[e.g.][]{Martizzi2016, Eckert2019, Ettori2022, Braspenning2024}, but without unbiased observables, i.e. gravitational lensing or kinematic information, accurately quantifying this bias on an individual cluster basis remains elusive. 

In this work, we will focus exclusively on predicting the NTP generated inside the cluster's core when subject to kinetic AGN feedback in the form of energetic jet outbursts. Whilst the pressure balance in the central region of clusters has less overall importance to the hydrostatic bias,\footnote{The AGN region occupies a small volume. It is the NTP and its radial gradient at the measured cluster radius, e.g. $r_{500}$, which determines the hydrostatic bias of the cluster mass enclosed within that radius \citep[see][]{Eckert2019}.} the balance that exists remains important for understanding how AGN feedback operates and interacts with the intracluster gas. Importantly, we will predict the NTP without assuming a fixed kinetic energy coupling from the AGN to the gas, to avoid tuning our model to any desired results. We will do this by modelling from first principles how an idealised thermodynamic system evolves when subject to some external energy injection rate. We will then compare the predicted NTP, when simulating a Perseus-like environment, with the \textit{Hitomi} results. 

More theoretical understanding in the area of kinetic AGN feedback will be pivotal for predicting and interpreting the observations currently being made by XRISM (X-Ray Imaging and Spectroscopy Mission), which, as the successor to \textit{Hitomi}, will resolve the NTP in the core of many more nearby clusters. Equally, these future observations will provide a comparison against current AGN feedback models. Strengthening these models will enable semi-analytic galaxy formation models and cosmological simulations that model AGN feedback to be placed on a more robust footing.

We address the question of AGN-induced NTP from a theoretical perspective in this paper, which is structured as follows. In Section \ref{Section 2}, we describe the model we use to evolve a jet outburst population, and in Section \ref{Section 3}, we show how the energy injected by each outburst can be coupled to the intracluster gas, producing some NTP fraction. These results are given in Section \ref{Section 4}, where we present the mean profiles of the induced NTP fractions. In Section \ref{Section 5}, we investigate the relationship between NTP and AGN jet activity, which we use to constrain the AGN jet activity of the Perseus cluster using its NTP observations. We quantify the sensitivity of our results to key model assumptions in Section \ref{Section 6}. Our conclusions are given in Section \ref{Section 7}. 

\section{THE JET EVOLUTION MODEL}\label{Section 2}

We describe our assumed cluster environment and analytical jet evolution model below. In Section \ref{Section 2.1}, we specify the gas and dark matter profiles, then, in Section \ref{Section 2.2}, we make use of observationally-inferred probability distributions to populate a sample of jet sources that we model within the cluster. In Section \ref{Section 2.3}, we outline the analytical framework for radially evolving each jet source in time within the cluster. Finally, in Section \ref{Section 2.4}, we remove all jet sources that we consider inconsistent with our modelling (e.g. jets that will not leave the galaxy). The remaining jet sources form the population of `outbursts' we consider in our results.

\renewcommand{\arraystretch}{1.5}
\begin{table*}
\begin{tabular}{ |p{2cm}||p{2.2cm}|p{3cm}|p{2.8cm}|p{2.2cm}|p{3cm}| }
\hline
\rowcolor{lightgray!30}
 &  \multicolumn{1}{|c|}{$\boldsymbol{c}$} & \multicolumn{1}{|c|}{$\boldsymbol{\alpha}$} & \multicolumn{1}{|c|}{$\boldsymbol{\eta}$} & \multicolumn{1}{|c|}{$\boldsymbol{d}$} & \multicolumn{1}{|c|}{$\boldsymbol{\varepsilon}$} \\
\hline\hline
\rowcolor{white} \cellcolor{lightgray!30} Definition: & \textit{Concentration} & \textit{Inner density slope of the dark matter profile} & \textit{Fraction of cosmological baryon content}  & \textit{Dilution} & \textit{Inner density slope of the intracluster gas profile} \\
\hline
\rowcolor{white} \cellcolor{lightgray!30} Assumed values: &  $ c=2.5$ & $\alpha \in [0, 1.5]$ & $\eta \in [0.6, 1]$ & $ d=1$ & $\varepsilon \in [0, 1]$\footnotemark \\
\hline
\end{tabular}
\caption{Summary of the five parameters in the analytical cluster model from \citetalias{Sullivan2024a} --- their symbol, definition and assumed values when $\Delta=500$.}
\label{Table - Environment parameters}
\end{table*}

\renewcommand{\arraystretch}{1.5}
\begin{table*}
\begin{tabular}{|p{2.8cm}|p{2.8cm}|p{2.8cm}|p{2.8cm}| }
\hline
\rowcolor{lightgray!30}
 \multicolumn{1}{|c|}{Virial mass} & \multicolumn{1}{|c|}{Virial radius} & \multicolumn{1}{|c|}{Virial temperature} & \multicolumn{1}{|c|}{Virial pressure} \\
\hline\hline
\rowcolor{white} $M_{500} = 10^{14.5}~\mathrm{M}_\odot $ & $r_{500} = 1.064 \times 10^3~\mathrm{kpc}$ & $T_{500} = 3.096 \times 10^7~\mathrm{K} $ & $p_{500} = 2.852 \times 10^{-13} ~\mathrm{Pa}$ \\
\hline
\end{tabular}
\caption{Summary of the virial parameters assumed for the galaxy cluster modelled in this study, when $\Delta=500$.}
\label{Table - Virial parameters}
\end{table*}

\subsection{The galaxy cluster environment}\label{Section 2.1}

\subsubsection{Galaxy clusters in virial equilibrium}\label{Section 2.1.1}

When describing galaxy clusters that are in virial equilibrium, the system's mass and physical radius are typically specified in terms of virial quantities. We use the definition
\begin{equation}\label{virial mass definition}
    M_\mathrm{vir} \equiv \frac{4}{3} \pi r_\mathrm{vir}^3 \Delta \rho_\mathrm{crit,0},
\end{equation}
whereby the virial mass, $M_\mathrm{vir}$, is defined as the mass enclosing an average density that is some overdensity, $\Delta$, times the present-day critical density, $\rho_\mathrm{crit,0}$, within the virial radius, $r_\mathrm{vir}$. \textcolor{black}{For the critical density, we take $\rho_\mathrm{crit, 0} = 1.8788\times 10^{-26}h^2$~kg~m$^{-3}$, and} assume a Hubble parameter of $h=0.6751$ \citep{Planck2016}.

In X-ray observations, $\Delta=500$ is the typical convention; this allows us to define the halo's corresponding virial mass, $M_{500}$, and virial radius, $r_{500}$. In this study, we will hereafter assume that the galaxy clusters we model are in virial equilibrium, and we will describe their physical properties by their virial parameters. 

\subsubsection{The dark matter and intracluster gas density profiles}\label{Section 2.1.2}

For our investigation, we will use the five-parameter analytical model proposed in \citet[][]{Sullivan2024a} (hereafter \citetalias{Sullivan2024a}) to model the structural composition of galaxy clusters. This model allows us to describe the radial density profiles of the dark matter, $\rho_\mathrm{dm}(s)$, and the intracluster gas, $\rho_\mathrm{gas}(s)$, each in scale-free form:
\begin{equation}\label{ideal physical baryonic halo profile - dark matter}
    \frac{\rho_\mathrm{dm} (s, c, \alpha, \eta)}{\Delta \rho_\mathrm{crit, 0}} = \frac{(1  - \eta f_\mathrm{b, cos}) u(c, \alpha)}{3s^\alpha (1 + cs)^{3-\alpha}},
\end{equation}
and
\begin{equation}\label{ideal physical baryonic halo profile - baryonic gas}
    \frac{\rho_\mathrm{gas}(s, c, \alpha, \eta, d, \varepsilon)}{\Delta \rho_\mathrm{crit,0}} = \frac{\eta f_\mathrm{b, cos}\mathcal{U}(c, \alpha, d, \varepsilon)}{3s^\varepsilon [1 + \mathcal{C}(c, \alpha, d, \varepsilon) s ]^{3-\varepsilon}},
\end{equation}
respectively. These profiles are given in terms of a scale-free dimensionless halocentric radius,
\begin{equation}\label{dimensionless halo radius}
    s \equiv \frac{r}{r_\mathrm{vir}}.
\end{equation}
\textcolor{black}{These profiles are constructed such that the mass enclosed at the virial radius is the virial mass, normalised to contain a baryon or gas mass fraction of $\eta f_\mathrm{b, cos}$ and a dark matter mass fraction of $1 - \eta f_\mathrm{b, cos}$.} We adopt a cosmological baryon fraction of $f_\mathrm{b, cos} = 0.158$ \citep{Planck2016}.

The five parameters that appear in these profiles are summarised in Table \ref{Table - Environment parameters}, along with their typical physical values when $\Delta=500$; the formal definitions for each of these parameters are the same as given in \citetalias{Sullivan2024a}. These parameters then specify the parameter functions in the above density profiles, which are
\begin{equation}\label{ideal physical halo concentration function}
    u(c, \alpha) \equiv \left[\int _0 ^1 \frac{s^{2-\alpha} \mathrm{d}s}{(1 + cs)^{3-\alpha}}\right]^{-1},
\end{equation}
\begin{equation}\label{baryon concentration parameter}
    \mathcal{C}(c, \alpha, d, \varepsilon) \equiv \frac{d(\alpha - \varepsilon) + c(3 - \varepsilon)}{3-\alpha},
\end{equation}
and
\begin{equation}\label{baryon concentration function}
	\mathcal{U}(c, \alpha, d, \varepsilon) \equiv \left[\int _0 ^{1} \frac{s^{2-\varepsilon}\mathrm{d}s}{[1 + \mathcal{C}(c, \alpha, d, \varepsilon) s]^{3-\varepsilon}} \right] ^{-1}.
\end{equation}

In our analysis that follows, we will consider the range of environmental parameters specified in Table \ref{Table - Environment parameters}. However, when showing figures we will specify the continuous parameters as those we consider to be a `typical cluster'. We refer to this as the cluster with $\alpha=1$, for a Navarro-Frenk-White (NFW) dark matter profile \citep[as informed by cosmological $N$-body simulations, see][]{NFW1995, NFW1996, NFW1997}; $\eta=0.8$, specifying the cluster to contain $80\%$ of the cosmological baryon content \citep[as a typical value constrained within $r_{500}$ for X-ray clusters, see, e.g.][]{Eckert2013, Pratt2023}; and $\varepsilon=0.5$, for a gas profile with a `weak cusp' \citep[as a typical value fit to X-ray cluster observations, see, e.g.][]{Lyskova2023, Pratt2023}. 

\subsubsection{Galaxy clusters in hydrostatic equilibrium}\label{Section 2.1.3}

\footnotetext{Revised to $\varepsilon \in [0, 0.5]$ in Section \ref{Section 4.2.1}.}

For a galaxy cluster described by these idealised profiles, the pressure profile that is in hydrostatic equilibrium with the gravitational potential of the cluster is (see \citetalias{Sullivan2024a})
\begin{equation}\label{ideal physical baryonic halo pressure}
	\frac{p_\mathrm{eq}(s, c, \alpha, \eta, d, \varepsilon)}{p_\mathrm{vir}} = \eta \mathcal{U}(c, \alpha, d, \varepsilon) \cdot \mathcal{I}(s, c, \alpha, \eta, d, \varepsilon),
\end{equation}
in terms of the integral function,
\begin{equation}\label{temperature integral function}
\begin{aligned}
    \mathcal{I}(s, c, \alpha, \eta, d, \varepsilon)&\equiv \int _s ^\infty   \frac{ \mathrm{d}s^\prime \Biggl\{  (1 - \eta f_\mathrm{b, cos})u(c, \alpha) \cdot \int _0 ^{s^\prime} \frac{s^\prime{}^\prime{} ^{2-\alpha} \mathrm{d}s^\prime{}^\prime}{(1 + cs^\prime{}^\prime )^{3-\alpha}}   }{s^\prime{}^{2 + \varepsilon} \left[1 + \mathcal{C}(c, \alpha, d, \varepsilon)s^\prime \right]^{3 - \varepsilon}} \\
    & \hspace{-15mm} + \quad \eta f_\mathrm{b, cos} \mathcal{U}(c, \alpha, d, \varepsilon) \cdot \int _0 ^{s^\prime} \frac{s^\prime{}^\prime{}^{2-\varepsilon} \mathrm{d}s^\prime{}^\prime}{\left[1 + \mathcal{C}(c, \alpha, d, \varepsilon)s^\prime{}^\prime \right]^{3-\varepsilon}} \Biggr\},
\end{aligned}
\end{equation}
and where
\begin{equation}\label{virial pressure definition}
    p_\mathrm{vir} \equiv \frac{k_\mathrm{B}T_\mathrm{vir}}{\mu m_\mathrm{p}} f_\mathrm{b, cos} \Delta \rho_\mathrm{crit,0}
\end{equation}
is the virial pressure, defined for some choice in $\Delta$. This virial parameter is specified by familiar physical quantities, as well as the Boltzmann constant, $k_\mathrm{B}$, and the proton mass, $m_\mathrm{p}$; \textcolor{black}{we assume a mean molecular weight of $\mu=0.60$.} The virial temperature is
\begin{equation}\label{virial temperature definition}
    T_\mathrm{vir} \equiv \frac{1}{3} \frac{\mu m_\mathrm{p}}{k_\mathrm{B}} \frac{GM_\mathrm{vir}}{r_\mathrm{vir}},
\end{equation}
defined in terms of the cluster's virial mass and virial radius, and \textcolor{black}{the gravitational constant, $G$.} When $\Delta=500$, these expressions allow us to define the virial parameters $T_{500}$ and $p_{500}$.

\subsubsection{Galaxy clusters in pristine equilibrium}\label{Section 2.1.4}

The equilibrium pressure solution, Equation \eqref{ideal physical baryonic halo pressure}, does not specify the thermal or NTP components of the gas pressure; we denote these pressures as $p_\mathrm{th}$ and $p_\mathrm{nt}$, respectively. If NTP was assumed to be absent from the cluster's state, i.e. $p_\mathrm{nt}=0$, then the equilibrium pressure would have to be purely thermal, i.e. $p_\mathrm{eq} = p_\mathrm{th}$. We will hereafter refer to this as the `pristine equilibrium' state of a cluster. In this assumption, the corresponding temperature of the gas is (see \citetalias{Sullivan2024a})
\begin{equation}\label{ideal physical baryonic halo temperature}
	\frac{T_\mathrm{eq}(s, c, \alpha, \eta, d, \varepsilon)}{T_\mathrm{vir}} = 3s^\varepsilon \left[1 + \mathcal{C}(c, \alpha, d, \varepsilon)s\right]^{3-\varepsilon} \cdot \mathcal{I}(s, c, \alpha, \eta, d, \varepsilon).
\end{equation}
If NTP was present at some location in the cluster, i.e. $p_\mathrm{nt} \ne 0$, assuming the cluster is still in hydrostatic equilibrium, the thermal pressure would be $p_\mathrm{th} < p_\mathrm{eq}$, thus reducing the gas temperature to be $T < T_\mathrm{eq}$, when assuming the gas density remains unchanged. If the amount of NTP is small, and thus these deviations are small, then the pristine equilibrium profiles provide a useful approximation. We will make use of these profiles throughout our study.

All of these profiles above are scale-free in the mass regime of galaxy clusters. However, for our analysis that follows, we must specify a particular mass scale, and thus fix the cluster's virial parameters. To be representative of a typical large galaxy cluster, we assume a cluster mass of $M_{500} = 10^{14.5}$~$\mathrm{M}_\odot$; this fixes the cluster's remaining virial parameters to be those given in Table \ref{Table - Virial parameters}. 

\subsubsection{The volumetric cooling rate of the galaxy cluster}\label{Section 2.1.5}

In maintenance-mode feedback, the AGN heating is expected to balance the background cooling of the intracluster gas. In this work, we will model the cluster's volumetric cooling rate as \citep[e.g.][]{Benson2010}
\begin{equation}\label{cooling rate}
    C(r, T) = n_\mathrm{H}^2(r) \Lambda (T),
\end{equation}
in terms of the number density profile of hydrogen gas, $n_\mathrm{H}(r)$, and the cooling function, $\Lambda(T)$, as a function of the cluster's temperature, $T$. To calculate this cooling rate, we assume that
\begin{equation}\label{hydrogen number density}
    n_\mathrm{H}(r) \simeq \frac{4}{9} \frac{\rho_\mathrm{gas}(r)}{\mu m_\mathrm{p}},
\end{equation}
where the gas density is modelled by Equation \eqref{ideal physical baryonic halo profile - baryonic gas}, and where the factor of $4/9$ is expected for a fully-ionised gas component, given the primordial elemental abundances \citep{Peebles1966, Schramm1977}. For the cooling function, we will determine its value at each cluster radius using the cooling function in \citet{SutherlandDopita1993}, interpolated over some radial temperature profile. If pristine equilibrium is assumed, the volumetric cooling rate will be approximated as
\begin{equation}\label{cooling rate - pristine equilibrium}
    C(r, T) \simeq C_\mathrm{eq}(r) = n_\mathrm{H}^2(r) \Lambda (T_\mathrm{eq}(r)),
\end{equation}
when interpolating over the pristine equilibrium temperature profile, Equation \eqref{ideal physical baryonic halo temperature}.

\subsection{Populating a sample of jet sources}\label{Section 2.2}

\subsubsection{Radio source populations}\label{Section 2.2.1}

\textcolor{black}{To model a realistic population of jet outbursts within the cluster, we make use of recent constraints on the properties of radio source populations. Under reasonable model assumptions, there exist approximate non-linear scaling relations between radio luminosity, environmental properties and jet power \citep[e.g.][]{Kaiser1997, Willott1999, ShabalaGodfrey2013}; this has been shown to hold for the majority of the lifetimes of long-lived sources. This allows the jet power of sources to be calculated from their observed radio luminosity.}

\textcolor{black}{\citet{KaiserBest2007} used this approach to show that the shape of the observed low-redshift radio luminosity function (RLF) can be explained by a broken power-law in jet powers, with a shallower slope for lower-powered sources compared to more powerful ones. This broken power-law in the RLF suggests that the probability distribution of jet powers, $P(Q_\mathrm{jet})$, can also be explained by a broken power-law, due to the known correlation between jet power and luminosity (albeit with many confounding variables).}

\textcolor{black}{For jet ages, their distribution will be related to the timescales regulating accretion. Whilst the specific mode of accretion does not need to be assumed to fit this distribution, a power-law of the form $P(t_\mathrm{jet}) \propto t_\mathrm{jet}^{-1}$ is expected if feedback regulation proceeds as pink noise. This is expected for Chaotic Cold Accretion (CCA), whereby cold gas condenses non-linearly into multiphase filaments that are accreted chaotically onto the black-hole \citep{Gaspari2017, Shabala2020, McKinley2022}. Other types of accretion are possible, e.g. hot accretion from the halo, but may not be associated with a pink noise distribution.}

\textcolor{black}{Over the past few years, the application of semi-analytic models \citep[e.g. Radio AGN in Semi-analytic Environments (RAiSE), ][]{TurnerShabala2015, Turner2023} to accurately determine the physical properties of observed radio populations has allowed stronger constraints to be placed on these distributions. In general, probability distributions of the form}
\begin{equation}\label{jet power distribution}
    P(Q_\mathrm{jet}) \propto Q_\mathrm{jet}^{-s_{Q}}, \quad Q_\mathrm{min} \leq Q_\mathrm{jet} \leq Q_\mathrm{max},
\end{equation}
\textcolor{black}{and}
\begin{equation}\label{jet age distribution}
    P(t_\mathrm{jet}) \propto t_\mathrm{jet}^{-s_{t}}, \quad t_\mathrm{min} \leq t_\mathrm{jet} \leq t_\mathrm{max},
\end{equation}
\textcolor{black}{have been fit to a range of observationally-inferred jet powers and ages, constraining their power-law slopes, $s_{Q}$ and $s_{t}$. Below, we refer to the results from \citet{Shabala2020} and \citet{Quici2024}, which both apply this modelling technique; we use their results to inform the probability distributions we use to populate the jet outbursts we model inside the cluster.}

\subsubsection{The jet power distribution}\label{Section 2.2.2}

In terms of their constraints on the distribution of jet powers, \citet{Shabala2020} suggest a best-fitting slope of $s_Q = 1$, whereas \citet{Quici2024} suggest a steeper slope of $s_Q = 1.5$. Comparing the observed populations of these studies, \citet{Quici2024} use a larger proportion of higher-powered radio sources than those in \citet{Shabala2020}. These differences can be interpreted as due to the break expected from the RLF \citep{KaiserBest2007}. 

As such, we model the jet power distribution as a broken power-law:
\begin{equation}\label{broken power law jet power distribution}
    P(Q_\mathrm{jet}) \propto 
    \begin{cases}
        Q_\mathrm{jet}^{-s_{Q,\mathrm{low}}}, \quad Q_\mathrm{min} \leq Q_\mathrm{jet} \leq Q_\mathrm{break}, \\
        Q_\mathrm{jet}^{-s_{Q,\mathrm{high}}}, \quad Q_\mathrm{break} < Q_\mathrm{jet} \leq Q_\mathrm{max},
    \end{cases}
\end{equation}
with $s_{Q,\mathrm{low}} = 1$ \citep[as in][]{Shabala2020} and $s_{Q,\mathrm{high}} = 1.5$ \citep[as in][]{Quici2024}, and continuity ensued at the break point. To specify the location of this break, we refer to \citet{Hardcastle2019}; they find that the peak jet power contribution to the jet kinematic luminosity function occurs at $Q_\mathrm{jet} = 10^{38}$~W, with the contribution falling off at higher-powered sources. We take this as the jet power of the break point, $Q_\mathrm{break}$.

To fully describe this probability distribution, we must choose the minimum, $Q_\mathrm{min}$, and maximum, $Q_\mathrm{max}$, jet powers to sample from. We take $Q_\mathrm{max} = 10^{40}$~W, consistent with the highest jet powers catalogued in radio AGN surveys. We note that this choice is not important, as the probability of such high-powered jets tapers off dramatically under a power-law slope of $s_{Q,\mathrm{high}}=1.5$. We constrain the value of $Q_\mathrm{min}$ in Section \ref{Section 2.4.3} below, when considering the potential confinement of jets in the central galaxy. 

\subsubsection{The jet age distribution}\label{Section 2.2.3}

For the distribution in jet ages, both \citet{Shabala2020} and \citet{Quici2024} suggest a power-law slope of $s_t = 1$, \textcolor{black}{i.e. a pink noise distribution, favouring CCA.} We adopt this value in our study. When sampling this distribution, we take $t_\mathrm{min}=1$~Myr as the minimum jet age, noting that this choice is trivial, as we will constrain the lifetime of each jet when considering the presence of a central galaxy in Section \ref{Section 2.4.2}. For the maximum jet age, we take $t_\mathrm{max} =1$~Gyr. 

\subsubsection{Creating the jet sample}\label{Section 2.2.4}

We generate a random sample of $N$ jet sources by independently sampling their jet power and age from the power-law distributions in Equations \eqref{broken power law jet power distribution} and \eqref{jet age distribution}, respectively. Whilst higher-powered radio sources appear to be longer-lived \citep{Hardcastle2019}, we expect this independent sampling to be valid for the dominant lower-powered jet population.

\subsection{A Fanaroff-Riley type I jet model}\label{Section 2.3}

\subsubsection{Two-phase jet-lobe expansion}\label{Section 2.3.1}

\textcolor{black}{Radio sources display a `core-jet-lobe' morphology. Historically, the Fanaroff-Riley classification \citep{FanaroffRiley} has been used to distinguish between sources that appear brighter in the jet region near the cluster's centre (FR-I sources), and those that appear brighter at the edge of the jet's lobes (FR-II sources). In this work, we assume that each source has an FR-I morphology. These source types are typical of lower-powered jet outbursts \citep[e.g.][]{Best2009} which dominate source counts. This will be reflected in our power-law distribution, which favours lower-powered sources. We discuss later (in Section \ref{Section 6.1.1}) that separately modelling higher-powered jet outbursts with an FR-II model would not strongly impact our results.} 

For FR-I sources, the jet first evolves in a `ballistic phase', as the jet propagates at high pressure through a constant cross-section of the cluster. During this phase, the jet's length is well-modelled to evolve in time as $R(t) \propto t^{2/(4-\varepsilon)}$ \citep[e.g.][]{Turner2023}, in terms of the negative logarithmic inner density slope of the gas profile, $\varepsilon$. At some point during the outburst, the jet begins to `flare': expanding into a lobe, filling a larger volume of the cluster, and becoming fainter. During this `flaring jet phase', the jet's length \textcolor{black}{evolves} more weakly in time, as $R(t) \propto t^{1/(3-\varepsilon)}$ \citep[e.g.][]{LuoSadler2010}.

In our model, we assume that the transition between these evolutionary regimes, referred to as the `flaring point', occurs at the time, $t=t_\mathrm{eq}$, in which the \textcolor{black}{pressure of the jet falls to the ambient pressure of its host cluster, establishing pressure equilibrium between the jet and the environment.} \textcolor{black}{We note that various other transition points have been postulated in the literature. These include the formation of a lobe when the mass swept up by the jet exceeds that launched by the jet \citep{KaiserBest2007}, and more detailed considerations of the jet's pressure that are broadly consistent with our assumption \citep{Krause2012}.}

Our model for the jet's evolution is summarised as follows:
\begin{enumerate}
    \item The ballistic phase, $R(t) \propto t^{2/(4-\varepsilon)}$, when $t\leq t_\mathrm{eq}$.
    \item The flaring jet phase, $R(t) \propto t^{1/(3-\varepsilon)}$, when $t> t_\mathrm{eq}$.
\end{enumerate}
The continuity between these evolutionary phases will be ensured in our model, as discussed later in this section. In the model below, we will evolve each jet's length, $R(t)$, over time, $t$, as a single-side axisymmetric jet lobe, which is duplicated to form an identical pair of axisymmetric jets propagating in opposite directions from the black-hole at the cluster's centre. 

\subsubsection{The ballistic phase}\label{Section 2.3.2}

During the ballistic phase, we assume that the jet subtends a constant initial half-opening angle at the black-hole, $\theta_\mathrm{jet}$, which fills the solid angle
\begin{equation}
    \Omega = 2\pi \left[1 - \cos \theta_\mathrm{jet}\right].
\end{equation}
We use the general expression for the evolution of the jet's length with time from \citet{Turner2023}:
\begin{equation}\label{jet phase}
    R(t\leq t_\mathrm{eq}) = \left[\frac{(4 - \varepsilon)t}{2}\right]^{2/(4-\varepsilon)} \left[\frac{\kappa_1Q_\mathrm{jet}}{\Omega c_\gamma \rho_\mathrm{gas}(r_0) r_0^\varepsilon  }\right]^{1/(4-\varepsilon)},
\end{equation}
where $c_\gamma$\footnote{The speed of light is denoted with a subscript, as $c_\gamma$, to avoid confusion with the halo's concentration, denoted as $c$.} is the speed of light, \textcolor{black}{and $1 \leq \kappa_1 < 2$ is a dimensionless constant, which we set to unity.} Equation \eqref{jet phase} is parameterised by the solid angle, $\Omega$, the single jet's power, $Q_\mathrm{jet}$, and the surrounding gas density profile, assumed to be a power-law, i.e. $\rho_\mathrm{gas}(r)\propto r^{-\varepsilon}$, where we use the form in Equation \eqref{ideal physical baryonic halo profile - baryonic gas}, which follows this power-law in the core. \textcolor{black}{This gas density profile is then evaluated at some small halocentric radius where the power-law holds; we use $r_0 = 0.01 r_{500}$.}

The jet's pressure during this ballistic expansion is \citep{Turner2023}
\begin{equation}\label{jet pressure}
    p_\mathrm{jet}(t \leq t_\mathrm{eq}) = \frac{\kappa_1 Q_\mathrm{jet}}{\Omega R^2(t)c_\gamma },
\end{equation}
in terms of the jet's length, $R(t)$, in Equation \eqref{jet phase}. We can reasonably assume that the ambient pressure of the environment that the jet propagates into is close to being in hydrostatic equilibrium with the galaxy cluster's gravitational potential. The time by which the \textcolor{black}{pressure of the jet, Equation \eqref{jet pressure}, has fallen to the same value as} the cluster's hydrostatic equilibrium pressure, i.e. when $p_\mathrm{jet}(t=t_\mathrm{eq}) = p_\mathrm{eq}(r)$, is
\begin{equation}\label{equilibrium time - general}
    t_\mathrm{eq} = \frac{2}{(4-\varepsilon)}  \left\{\left[\frac{\kappa_1 Q_\mathrm{jet}}{\Omega c_\gamma }\right]^{(2-\varepsilon + \varepsilon^\prime)}\frac{[\rho_\mathrm{gas}(r_0) r_0 ^\varepsilon]^{(2-\varepsilon^\prime)}}{[p_\mathrm{eq}(r_0)r_0^{\varepsilon^\prime}]^{(4-\varepsilon)}}\right\}^{1/(4-2\varepsilon^\prime)},
\end{equation}
where we assume this hydrostatic equilibrium pressure is also a power-law, i.e. $p_\mathrm{eq}(r) \propto r^{-\varepsilon^\prime}$, for some different inner slope, $\varepsilon^\prime$. For pressure profiles of the form in Equation \eqref{ideal physical baryonic halo pressure}, this inner slope is $\varepsilon^\prime \approx 0$ \textcolor{black}{at $r_0 = 0.01r_{500}$} for most\footnote{Strictly, this inner slope, $\varepsilon^\prime$, is non-zero for some cuspy dark matter halos, when $\alpha=1.5$, and some cuspy gas profiles, when $\varepsilon=1$. However, these environments are either not considered or not important for our results (see Section \ref{Section 4.2.1}) and so we do not consider these mild corrections here.} cluster environments we consider in Table \ref{Table - Environment parameters} (see \citetalias{Sullivan2024a}); as such, we assume a constant-pressure gas environment, $p_\mathrm{eq}(r_0)$, by evaluating these profiles in the core. This allows us to simplify Equation \eqref{equilibrium time - general} to
\begin{equation}\label{equilibrium time}
    t_\mathrm{eq} = \frac{2}{(4-\varepsilon)}  \left[\frac{\kappa_1 Q_\mathrm{jet}}{\Omega c_\gamma }\right]^{(2-\varepsilon)/4} \frac{[\rho_\mathrm{gas}(r_0) r_0 ^\varepsilon]^{1/2}}{[p_\mathrm{eq}(r_0)]^{(4-\varepsilon)/4}}.
\end{equation}
We refer to this time as the equilibrium time. We define the jet's length at this time as the equilibrium jet length, which is
\begin{equation}
    R_\mathrm{eq} \equiv R(t=t_\mathrm{eq})= \left[\frac{\kappa_1 Q_\mathrm{jet}}{\Omega c_\gamma p_\mathrm{eq}(r_0)}\right]^{1/2}.
\end{equation}
\textcolor{black}{In a typical cluster environment, our model predicts equilibrium times of $\sim$ a few Myr, and equilibrium jet lengths of $\sim$ a few kpc (the distributions of these properties are shown in Section \ref{Section 4.1.1}).}

\begin{figure}
    \centering
    \includegraphics[width=0.5\textwidth]{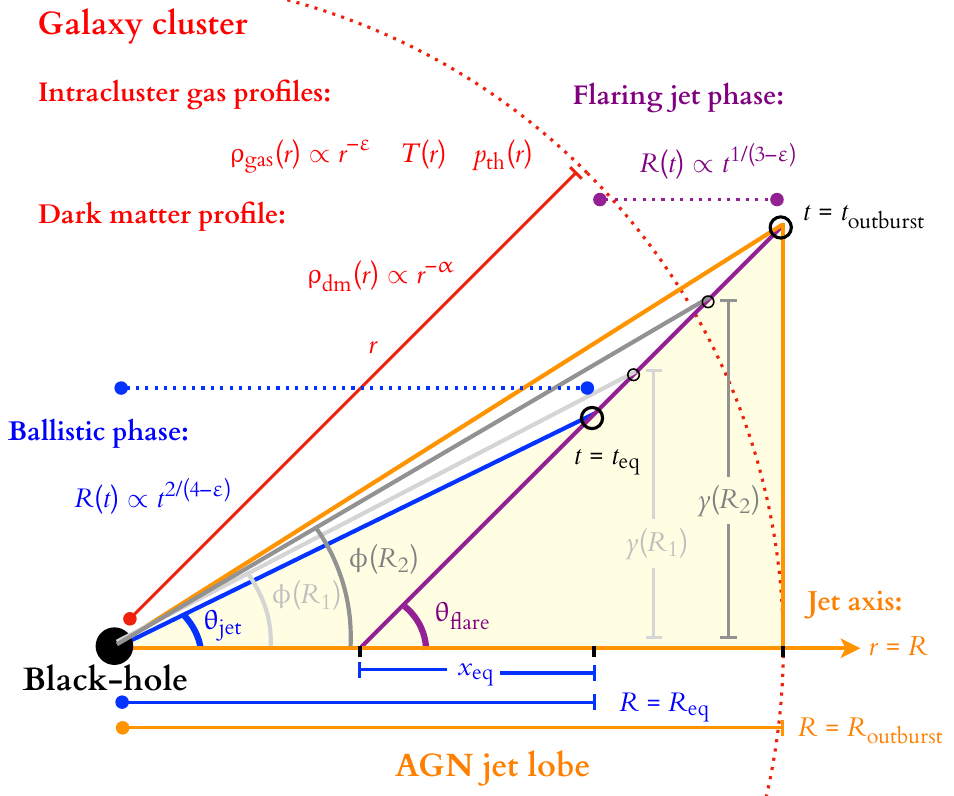}
    \caption{Our FR-I jet-lobe model embedded within a galaxy cluster.}
    \label{FR-I jet diagram}
\end{figure}

\subsubsection{The flaring jet phase}\label{Section 2.3.3}

Once the jet pressure reaches that of the cluster environment, the jet will begin to flare. This transition is illustrated in Figure \ref{FR-I jet diagram}, where beyond the flaring point, $t=t_\mathrm{eq}$, \textcolor{black}{the half-opening angle subtended by the jet at the black-hole begins increasing as a function of the jet's length; this angle is $\phi(R)$, which in the diagram is evaluated at two separate points along the jet axis. This flaring region can be parametrised by assuming a constant half-opening angle subtended by the jet at the flaring point; this is $\theta_\mathrm{flare}$.}

To model the evolution of the jet's length beyond this flaring point, we take the expression for the self-similar pressure-limited jet expansion from \citet{LuoSadler2010}: 
\begin{equation}
    R(t) = R_0 \left(\frac{t}{t_0}\right)^{1/(3-\varepsilon)},
\end{equation}
which is equivalent to the subsonic jet expansion in \citet{TurnerShabala2015}, but for our case we must take into account the jet's previous expansion history. For our evolutionary model, this is accounted for in the expression:
\begin{equation}\label{flaring jet phase}
    R(t>t_\mathrm{eq}) = \left(R_\mathrm{eq} - x_\mathrm{eq}\right) + x_\mathrm{eq} \left[\frac{(t-t_\mathrm{eq} + \Delta t_\mathrm{flare})}{\Delta t_\mathrm{flare}}\right]^{1/(3-\varepsilon)},
\end{equation}
where 
\begin{equation}
    x_\mathrm{eq} = \frac{R_\mathrm{eq} \tan \theta_\mathrm{jet}}{\tan \theta _\mathrm{flare}}
\end{equation}
gives the length offset in this transition, and $\Delta t_\mathrm{flare}$ (which will always be positive) is a time-offset term introduced to ensure continuity. 

To connect the two evolutionary phases, we demand that the jet's velocity, i.e. the time-derivative of the jet's length, $\dot{R}(t)$, is continuous at the flaring point: $\dot{R}(t\leq t_\mathrm{eq}) = \dot{R}(t> t_\mathrm{eq})$ at $t=t_\mathrm{eq}$. This constrains the flaring jet time-offset as
\begin{equation}\label{tfj}
    \Delta t_\mathrm{flare} = \frac{1}{2}\frac{(4-\varepsilon)}{(3-\varepsilon)} \frac{x_\mathrm{eq}}{R_\mathrm{eq}}t_\mathrm{eq}.
\end{equation}

\subsubsection{The filling factor of the jet}\label{Section 2.3.4}

We now consider how jets described by this evolutionary model propagate through the volume of the galaxy cluster. In Figure \ref{FR-I jet diagram}, the half-angle subtended by the jet at the black-hole is given by $\phi(R)$, as a function of the jet's length, $R$, along its propagation axis. Before the jet flares, this angle is simply the initial half-opening angle of the jet, i.e. $\phi(R\leq R_\mathrm{eq}) = \theta_\mathrm{jet}$; beyond this flaring point, this angle increases as
\begin{equation}
    \phi (R>R_\mathrm{eq}) = \arccos \left[\frac{R}{\sqrt{R^2 + y^2(R)}}\right],
\end{equation}
where
\begin{equation}
    y(R) = \left[R(t) - R_\mathrm{eq} + x_\mathrm{eq}\right]\tan\theta_\mathrm{flare}
\end{equation}
is the height of the jet lobe above the jet axis (see Figure \ref{FR-I jet diagram}). At all points along the jet axis, this half-opening angle, $\phi(R)$, determines the solid angle that the jet fills at that jet length, which is
\begin{equation}
    \Phi(R) = 2\pi \left[1 - \cos \phi (R)\right].
\end{equation}

We can parameterise the propagation of the jet through the cluster by its filling factor, $f$. \textcolor{black}{We take this filling factor to be the fraction of the cluster's surface area occupied by the jet lobes, at any jet length. This ensures that the radial filling factor, $f(r)$, effectively parameterises the volume occupation of the jet lobes inside infinitesimally thin spherical shells.} For a spherical cluster with a pair of identical axisymmetric jet lobes launched by the black-hole at the cluster's centre, this filling factor is 
\begin{equation}
    f(r \leq R_\mathrm{outburst}) = \frac{2\Phi(R=r)}{4\pi} = \left[1 - \cos \phi (r)\right],
\end{equation}
where the jet's length, $R$, is aligned with the cluster's halocentric radius, $r$, and where we denote $R_\mathrm{outburst} \equiv R(t=t_\mathrm{outburst})$ as the final length of each jet lobe at the end of the outburst. The filling factor beyond this point is zero, i.e. $f(r > R_\mathrm{outburst})=0$.

\subsubsection{\textcolor{black}{The half-opening angles of the jet}}\label{Section 2.3.5}

\textcolor{black}{In this evolutionary model, we must specify the jet's half-opening angles, $\theta_\mathrm{jet}$ and $\theta_\mathrm{flare}$ (defined in Figure \ref{FR-I jet diagram}). We briefly motivate observationally-inferred values for these parameters below, albeit we note there are limited measurements available.}

\textcolor{black}{M87 is the most studied jet: it has a very broad half-opening angle of $\sim 30^\circ$ on the sub-$\mathrm{pc}$ scale, which is strongly collimated over $\sim$~10~pc and reduced to $\lesssim 5^\circ$ on the kpc-scale \citep{Junor1999}. Simulations typically launch jets with an initial half-opening angle of $\lesssim 10^\circ$ \citep[e.g.][]{Young2024} to model the already collimated jet. Once the jet begins to flare, this opening angle increases; \citet{LaingBridle2014} studied this increase for 10 FR-I sources, where they found a median half-opening angle of $\sim 16^\circ$ in the region before recollimation occurs around $5-15$~kpc. In our work, we assume half-opening angles of $\theta_\mathrm{jet} = 8.5^\circ$ and $\theta_\mathrm{flare}=15^\circ$ in the ballistic and flaring regions of the jet, respectively, broadly consistent with these measurements.}

\subsection{Removing select jet sources}\label{Section 2.4}

\subsubsection{Modelling the brightest cluster galaxy (BCG)}\label{Section 2.4.1}

In this evolutionary model, so far we have neglected the presence of the cluster's central galaxy, also called the `brightest cluster galaxy' (BCG). In real systems, lower-powered (or short-lived) jets will not make it out of the cluster's BCG due to its high gas and stellar density. These sources will thus never produce an outburst on scales relevant to this work. This imposes a constraint on the minimum jet power, $Q_\mathrm{min}$, which can be used to inform the distribution of jet powers in Equation \eqref{broken power law jet power distribution}. This calibration will also constrain each source's age, as the jet will spend some amount of time first escaping the BCG before reaching the cluster environment.

To model the gas density in the BCG, we assume a density profile with the same inner slope, $\varepsilon$, as the background intracluster gas profile. We note that if these gas density profiles were modelled with separate inner slopes, the change to our results would be similar to adjusting the gas mass of the BCG, which we investigate in Section \ref{Section 6.1.2}. We specify the BCG's gas density profile as
\begin{equation}\label{bcg gas density}
    \rho_\mathrm{gas, \, bcg}(r) = \frac{(3-\varepsilon)M_\mathrm{gas}}{4 \pi r_\mathrm{gas}^{3-\varepsilon}} \frac{1}{r^\varepsilon},
\end{equation}
where $M_\mathrm{gas}$ is the BCG mass enclosed within some radius, $r_\mathrm{gas}$. 

We define the escape time, $t_\mathrm{esc}$, as the time it takes for a given jet to reach the outskirts of the BCG, i.e. when $R(t=t_\mathrm{esc}) = r_\mathrm{gas}$. This time will depend on the BCG's gas density, the jet power of each source, and the cross-section of the jet. As the jet will be collimated in the galaxy, we model its propagation by some effective solid angle, which we denote $\Omega_\mathrm{eff}$. By using Equation \eqref{jet phase} to model the jet's evolution in the galaxy, we must then solve the condition
\begin{equation}
    r_\mathrm{gas} = \left[\frac{(4 - \varepsilon)t_\mathrm{esc}}{2}\right]^{2/(4-\varepsilon)} \left[\frac{\kappa_1Q_\mathrm{jet}}{\Omega_\mathrm{eff} c_\gamma \rho_\mathrm{gas, \, bcg}(r_0) r_0^\varepsilon  }\right]^{1/(4-\varepsilon)}
\end{equation}
for the escape time, $t_\mathrm{esc}$. In terms of the properties of the BCG, using Equation \eqref{bcg gas density}, the escape time for a given source is then
\begin{equation}\label{escape time}
    t_\mathrm{esc} = \frac{1}{(4 - \varepsilon)} \left[\frac{(3-\varepsilon)\Omega_\mathrm{eff} c_\gamma M_\mathrm{gas} r_\mathrm{gas} }{\pi\kappa_1 Q_\mathrm{jet}}\right]^{1/2}.
\end{equation}

To simplify Equation \eqref{escape time}, we calibrate to the time it takes a simulated jet outburst to escape a BCG. We use the hydrodynamic jet simulations from \citet{Young2024}, which give the time it takes a $Q_\mathrm{jet}=10^{37}$~W jet outburst to escape three realisations of a clumpy interstellar medium for a BCG with known radius and central density. These outbursts take on average 1.27~Myr to reach a radius of 2.5~kpc, with this radius corresponding to the edge of the BCG, and enclosing a gas mass of $2.67\times 10^{10}$~$ \mathrm{M}_\odot$, in a predominantly constant gas density, i.e. $\varepsilon\approx0$, environment. This allows us to scale Equation \eqref{escape time} as
\begin{equation}\label{calibrated escape time}
\begin{split}
    \frac{t_\mathrm{esc}}{1.27~\mathrm{Myr}} &= \frac{4}{(4-\varepsilon)} \left[\frac{(3-\varepsilon)}{3}\right]^{1/2} \left[\frac{10^{37}~\mathrm{W}}{Q_\mathrm{jet}}\right]^{1/2} \\
    & \times \left[\frac{M_\mathrm{gas}}{2.67\times 10^{10} ~\mathrm{M}_\odot}\right]^{1/2} \left[\frac{r_\mathrm{gas}}{2.5 ~\mathrm{kpc}}\right]^{1/2},
\end{split}
\end{equation}
where the dependence on the effective solid angle, $\Omega_\mathrm{eff}$, is mitigated by this calibration; we assume henceforth that this parameter is accurately modelled in the simulation. 

For our calculation, we will assume that the BCG inside our galaxy cluster is exactly that modelled in \citet{Young2024}, i.e. with $M_\mathrm{gas} = 2.67\times 10^{10}$~$\mathrm{M}_\odot$ and $r_\mathrm{gas}=2.5$~kpc. Typical BCG gas masses of $\sim$ a few $\times~10^{10}$~$\mathrm{M}_\odot$ have been measured in nearby clusters \citep[e.g.][]{Dunne2021}; \textcolor{black}{their radius will vary substantially for different galaxy types, with ellipticals typically extending further along the jet axis, and spirals a lesser distance.} \textcolor{black}{The fact that real BCGs will vary substantially in their morphology is only important to our calibration if this changes the amount of each jet's available lifetime (and energy) that is spent in outburst.} If this modelled BCG changed in $M_\mathrm{gas}$ or $r_\mathrm{gas}$ by some multiplicative factor of $A$ to account for this variation, a multiplicative factor of $\sqrt{A}$ would propagate into Equation \eqref{calibrated escape time}. The effects of this change are investigated in Section \ref{Section 6.1.2}, where we show the impact on our results to be very minimal. 

\textcolor{black}{Due to the variance of possible BCG morphologies, we have not included the BCG in the cluster's gravitational potential in this work. We only consider the BCG for this escape time calibration. If the BCG mass was included in the gravitational potential, we expect that its contribution would be less than the variation that occurs when varying the dark matter halo profile. At small radius, the enclosed mass can range between $M(<2.5~\mathrm{kpc})\sim 10^{8}-10^{11}$~$\mathrm{M}_\odot$ and $M(<10~\mathrm{kpc})\sim 10^{9}-10^{12}$~$\mathrm{M}_\odot$ when the inner slope of the dark matter profile is steepened from $\alpha =0$ to $\alpha=1.5$, which is accounted for in our results (see Section \ref{Section 4.2.1}). As such, it is only the extent of the BCG into the cluster, i.e. the value of $r_\mathrm{gas}$, which will be relevant to our model; within this region, the jet will not interact with the cluster medium, and so will not induce NTP into the intracluster gas. This will mask or obscure our results within the extent of a particular cluster's BCG.}

\subsubsection{Removing jets that do not produce outbursts}\label{Section 2.4.2}

Given that each jet will spend some amount of time, $t_\mathrm{esc}$, escaping the cluster's BCG, for each jet we sample we must now subtract this time from its sampled jet age, $t_\mathrm{jet}$, to deduce the jet's active or `outburst time',
\begin{equation}
    t_\mathrm{outburst} \equiv t_\mathrm{jet} - t_\mathrm{esc}.
\end{equation}
If the time it takes for a given jet to escape the BCG is larger than its sampled age, i.e. if $t_\mathrm{esc} > t_\mathrm{jet}$, then the source will never produce an outburst that directly impacts the cluster. We remove these sources from our sample. This allows us to form a subset of jets, $N_\mathrm{esc}$, which escape the BCG and produce outbursts relevant for this work. As this escape time depends on the jet's power, Equation \eqref{calibrated escape time}, this constraint is equivalent to a linear cut in the $Q_\mathrm{jet}$-$t_\mathrm{jet}$ plane, where all sources in our sample of jet power less than $Q_\mathrm{jet}(t_\mathrm{esc}=t_\mathrm{jet})$ for a given $t_\mathrm{jet}$ are removed from our sample. This circumvents the sensitivity of our model to the value of $t_\mathrm{min}$ in the probability distribution in Equation \eqref{jet age distribution}, as the minimum time a jet can spend in outburst can now be fixed in terms of its jet power.

\subsubsection{Constraining the minimum jet power}\label{Section 2.4.3}

We can now specify the global value of $Q_\mathrm{min}$ in the jet power probability distribution, Equation \eqref{broken power law jet power distribution}, as a limiting value of the linear cut in the $Q_\mathrm{jet}$-$t_\mathrm{jet}$ plane given by the BCG calibration. We take this to be $Q_\mathrm{min}\equiv Q_\mathrm{jet}(t_\mathrm{esc}=1$~Gyr), as the minimum jet power that escapes the BCG within 1~Gyr, consistent with our choice in $t_\mathrm{max}$. This gives the calibration
\begin{equation}\label{calibrated jet power}
\begin{split}
    \frac{Q_\mathrm{min}}{1.613\times 10^{31}~\mathrm{W}} &= \left[\frac{1~\mathrm{Gyr}}{t_\mathrm{esc}}\right]^2 \left[\frac{4}{(4-\varepsilon)} \right]^2 \frac{(3-\varepsilon)}{3}  \\
    & \times \left[\frac{M_\mathrm{gas}}{2.67\times 10^{10} ~\mathrm{M}_\odot}\right] \left[\frac{r_\mathrm{gas}}{2.5 ~\mathrm{kpc}}\right],
\end{split}
\end{equation}
with an absolute minimum jet power of $Q_\mathrm{min}\simeq 1.6\times 10^{31}$~W, which has a slight dependence on the slope, $\varepsilon$, of the gas profile. If the BCG we model changed by a factor of $A$ in $M_\mathrm{gas}$ or $r_\mathrm{gas}$, this would propagate a multiplicative factor of $A$ into Equation \eqref{calibrated jet power}. As before, these effects are minimal, as explored in more detail in Section \ref{Section 6.1.2}.

\subsubsection{Removing jets that never reach pressure equilibrium}\label{Section 2.4.4}

After this calibration to ensure that each jet escapes the BCG, the sample will be left with some sources that have very small outburst times that only just escape the galaxy. For these sources, the jet will never reach far enough out in the cluster to reach pressure equilibrium, i.e. $t_\mathrm{outburst} < t_\mathrm{eq}$, which implies that these sources will never enter a flaring jet phase, and so will never form a lobe. This is doubly problematic for our model, as i) the jet will never be in pressure equilibrium with the cluster during these sources' lifetimes, and ii) all of the jet's power will be deposited in a very small volume of the cluster, leading to unphysically high volumetric heating rates. For these reasons, we choose to remove these sources from our population. This leaves us with a reduced sample of jets, $N_\mathrm{flare} < N_\mathrm{esc}$, containing all jets that both escape the BCG \textit{and then} go on to flare. This forms the population we will consider for our results hereafter.

\section{MODELLING KINETIC AGN FEEDBACK}\label{Section 3}

Here we outline our approach to modelling the kinetic AGN feedback induced by our population of jet outbursts. In Section \ref{Section 3.1}, we introduce the AGN jet duty cycle and use this to characterise the heating rates of the jet outbursts. In Section \ref{Section 3.2}, we show how the intracluster gas can evolve when modelled as a thermodynamic system experiencing an external energy injection rate, with the kinetic energy coupling to the gas in each jet outburst determined implicitly in the model. In Section \ref{Section 3.3}, we describe how the NTP fraction can be calculated in this framework.

\subsection{Heating rates from AGN jet outbursts}\label{Section 3.1}

\subsubsection{The AGN jet duty cycle}\label{Section 3.1.1}

The timescale during which an AGN is active is typically quantified by the AGN jet duty cycle. For a given source, this is the fraction of the AGN lifecycle, $t_\mathrm{life}$, that the AGN jets are `switched on'. We define the AGN jet duty cycle in this study as
\begin{equation}\label{duty cycle}
    \delta \equiv \frac{t_\mathrm{on}}{t_\mathrm{life}}=\frac{t_\mathrm{on}}{t_\mathrm{on} + t_\mathrm{off}},
\end{equation}
where $t_\mathrm{on}$ and $t_\mathrm{off}$ give the average amounts of time that the AGN is in its `on' and `off' states. \textcolor{black}{This on/off classification corresponds to whether the AGN is radio-loud (i.e. is an active radio source). \textcolor{black}{We note that the definition of radio-loud can be ambiguous in the literature; we assume this corresponds to an intrinsic radio luminosity cut, consistent with most surveys of radio galaxies \citep[e.g.][]{Best2005, Sabater2019}.} This definition allows the AGN jet duty cycle to be estimated observationally from the fraction of the AGN population that appears radio-loud; this is called the radio-loud fraction.}

The AGN jet duty cycle is known to be a strong function of host galaxy properties \citep{Best2005, Shabala2008, Sabater2019}, with larger, more rapidly cooling structures requiring higher AGN jet duty cycles to maintain a heating-cooling balance \citep{Pope2012}. We consider the value of $\delta=10\%$, \textcolor{black}{which corresponds to the approximate percentage of radio-loud galaxies that are observed to have jets; this value is consistent with} the AGN activity inferred in observational surveys \citep[e.g.][]{Sun2015}. We will also consider higher values of $\delta=20\%$ and $\delta=30\%$, to consider the impact that increased AGN jet activity has on our results.

\textcolor{black}{Theory suggests that the AGN's central black-hole will always maintain a positive mass accretion rate, and therefore some amount of energy will always be available to sustain a jet, even if the jet does not escape the galaxy and produce a radio structure. In our model, accretion will be approximately constant during the outburst, dropping rapidly after the outburst ends, causing the jet to disrupt, and decoupling the state of the accretion disk from the extended outburst. This enables the AGN to become radio-quiet even whilst accretion continues; this mode of regulation follows the statistics of pink noise.}

\subsubsection{Accounting for compact sources in the AGN population}\label{Section 3.1.2}

\textcolor{black}{In our model, we only consider jet outbursts that are presently fuelling radio structures beyond the galaxy, which we calibrated to via a linear cut in the $Q_\mathrm{jet}$-$t_\mathrm{jet}$ plane (Sections \ref{Section 2.4.1}-\ref{Section 2.4.3}), i.e. these are the jets that escape the galaxy. However, there will be some fraction of radio-loud AGN that do not have radio structures extending beyond the host galaxy; these jets are confined to the galaxy. We refer to these as compact sources. Whilst it is extended sources that will produce kinetic feedback and interact with the cluster medium, it is important to note that compact sources still contribute to the radio-loud fraction. These sources thus need to be included in our population, given our definition of the AGN jet duty cycle.} 

We assume that compact sources comprise some fraction, $f_\mathrm{compact}$, of the total radio-loud AGN population. Approximately $30\%$ and $10\%$ of AGN sources are compact steep spectrum (CSS) and gigahertz peaked spectrum (GPS) sources, respectively \citep[e.g.][]{ODea1998}; we will assume $f_\mathrm{compact} = 0.4$. We can then introduce a population of
\begin{equation}\label{compact sources}
    N_\mathrm{compact} = \frac{f_\mathrm{compact} }{(1-f_\mathrm{compact})}N_\mathrm{flare}
\end{equation}
sources, which are confined to the galaxy, have no jet lobes and zero heating everywhere, and experience only the energy injection from the cluster's background cooling rate. \textcolor{black}{The inclusion of these sources in our sample, i.e. $N_\mathrm{agn} = N_\mathrm{flare} + N_\mathrm{compact}$, will be representative of the total radio-loud AGN population. This ensures that the AGN jet duty cycle defined in Equation \eqref{duty cycle} accounts for both extended and compact sources. We will use this total AGN population to calculate the mean profiles we give in our results.}

\renewcommand{\arraystretch}{1.5}
\begin{table*}
\begin{tabular}{ |p{2cm}||p{4cm}|p{4cm}|}
\hline
\rowcolor{lightgray!30}
 &  \multicolumn{1}{|c|}{$\boldsymbol{\theta}_\mathrm{jet}$} & \multicolumn{1}{|c|}{$\boldsymbol{\theta}_\mathrm{flare}$} \\
\hline\hline
\rowcolor{white} \cellcolor{lightgray!30} Definition: & \textit{Initial half-opening angle subtended by the jet at the black-hole} & \textit{Half-opening angle subtended by the jet at the flaring point}  \\
\hline
\rowcolor{white} \cellcolor{lightgray!30} Assumed values: &  $ \theta_\mathrm{jet} = 8.5^\circ $ & $\theta_\mathrm{flare} = 15^\circ$  \\
\hline
\end{tabular}

\vspace{2mm}

\renewcommand{\arraystretch}{1.5}
\begin{tabular}{ |p{2cm}||p{2.4cm}|p{2.8cm}|p{3.5cm}|p{3.5cm}|}
\hline
\rowcolor{lightgray!30}
 &  \multicolumn{1}{|c|}{$\boldsymbol{Q}_\mathrm{max}$} & \multicolumn{1}{|c|}{$\boldsymbol{Q}_\mathrm{break}$} & \multicolumn{1}{|c|}{$\boldsymbol{s}_{Q,\mathrm{low}}$}  & \multicolumn{1}{|c|}{$\boldsymbol{s}_{Q,\mathrm{high}}$} \\
\hline\hline
\rowcolor{white} \cellcolor{lightgray!30} Definition: & \textit{Maximum jet power} & \textit{Break in the jet power probability distribution} & \textit{Slope of the higher-powered jet power probability distribution} & \textit{Slope of the lower-powered jet power probability distribution} \\
\hline
\rowcolor{white} \cellcolor{lightgray!30} Assumed values: & $Q_\mathrm{max} = 10^{40}~\mathrm{W} $ & $Q_\mathrm{break} = 10^{38}~\mathrm{W} $  & $s_{Q,\mathrm{low}} = 1$ & $s_{Q,\mathrm{high}} = 1.5$ \\
\hline
\end{tabular}

\vspace{2mm}

\renewcommand{\arraystretch}{1.5}
\begin{tabular}{ |p{2cm}||p{2.4cm}|p{3.3cm}|p{3.6cm}|}
\hline
\rowcolor{lightgray!30}
& \multicolumn{1}{|c|}{$\boldsymbol{t}_\mathrm{max}$} & \multicolumn{1}{|c|}{$\boldsymbol{s}_t$} & \multicolumn{1}{|c|}{$\boldsymbol{\delta}$} \\
\hline\hline
\rowcolor{white} \cellcolor{lightgray!30} Definition: & \textit{Maximum jet age} & \textit{Slope of the jet age probability distribution} & \textit{The AGN jet duty cycle, defined in Equation \eqref{duty cycle}} \\
\hline
\rowcolor{white} \cellcolor{lightgray!30} Assumed values: & $t_\mathrm{max}= 1 \, \mathrm{Gyr} $ & $s_t = 1 $ & $\delta=10\%, 20\%$ and $30\%$ \\
\hline
\end{tabular}
\caption{Summary of the nine parameters in our FR-I jet heating model --- their symbol, definition and assumed values.}
\label{Table - Jet parameters}
\end{table*}

\subsubsection{The volumetric heating rate of the AGN jet}\label{Section 3.1.3}

For our extended source population, i.e. $N_\mathrm{flare}$, we now detail how we calculate the volumetric heating rate, $\mathcal{H}(r)$, for each jet outburst. We will assume that the total energy injected by two identical jet outbursts is spread uniformly within the total volume occupied by the jet lobes, $V_\mathrm{occupied}$, once they reach their final length, $R_\mathrm{outburst}$, when evolved in the FR-I model. This heating rate is then a constant value within the jet's volume, which along the jet axis is
\begin{equation}\label{jet heating rate}
    \mathcal{H}(r\leq R_\mathrm{outburst}) = \sqrt{\delta}\frac{2Q_\mathrm{jet}}{V_\mathrm{occupied}} = \frac{\sqrt{\delta} Q_\mathrm{jet}}{2\pi \int _0 ^{R_\mathrm{outburst}} r^2 f (r) \mathrm{d} r},
\end{equation}
and is zero beyond the outburst, i.e. $\mathcal{H}(r>R_\mathrm{outburst}) = 0$. 

In this expression, the $\sqrt{\delta}$ term appears due to the choice in modelling AGN feedback as a time-average of its `on' and `off' phases. When the AGN is switched off, the heating rate will be zero; when the AGN is switched on, the heating rate will be $\mathcal{H}_\mathrm{on} \equiv 2Q_\mathrm{jet}/V_\mathrm{occupied}$. As the induced NTP is proportional to the square of the AGN's volumetric heating rate \textcolor{black}{(this is shown explicitly in Section \ref{Section 3.3.3})}, the time-averaged NTP over the total AGN lifecycle is proportional to $(1-\delta)\times 0^2 + \delta \times \mathcal{H}_\mathrm{on}^2$; this implies that the time-averaged NTP depends on the effective volumetric heating rate, $\sqrt{\delta}\mathcal{H}_\mathrm{on}$. This is the profile given in Equation \eqref{jet heating rate}.

We justify modelling time-averaged AGN feedback as it allows us to account for the longer-lived energetics in the cluster core, instead of modelling the `on' and `off' feedback regimes separately. We also note that the timescale of a typical outburst will be shorter than the cluster's dynamical timescale. Long-lived gas motion is expected; turbulence will arise both in the active phase at the flaring point and during the quiescent phase due to buoyancy and fluid instabilities as the gas returns to hydrostatic equilibrium. These longer-lived energetics have been observed, where heating times of $\simeq 60\%$ and potentially up to $\simeq 100\%$ of the AGN lifecycle have been suggested, inferred from bubbles in the core that continue injecting power to the surrounding gas following outbursts \citep{Birzan2012}. 

\subsubsection{\textcolor{black}{Considerations for the AGN heating rate}}\label{Section 3.1.4}

\textcolor{black}{In the model described above, we only consider AGN heating to occur inside the volume filled by the jet lobes, as prescribed by the jet filling factor, $f(r)$, which will be parameterised by the assumed half-opening angles, $\theta_\mathrm{jet}$ and $\theta_\mathrm{flare}$. In real clusters, AGN jets are known to generate processes that occur outside the volume occupied by the jets. These include bow shocks, which are a type of shock wave that can occur when the jet's velocity becomes supersonic. Bow shocks have been observationally identified in the presence of jets \citep[e.g.][]{Motta2025}, and where they occur, they are expected to contribute energy to the surrounding medium.}

\textcolor{black}{To investigate the frequency at which jet-driven shock waves could be expected to occur in our model, we calculate the jet velocities for a sample of jets and compare these values to the sound speed of the cluster environment at all times during each outburst. Inside a typical cluster environment, we find that all jets are launched with supersonic velocities, but rapidly each jet's Mach value falls, and before the flaring point, all jets are subsonic, remaining so for their remaining lifetime. On average, over our population, we find that jets are subsonic for $\gtrsim 90\%$ of their outburst time. This implies that shocks will not be a major heating consideration for our model, and as such, we do not model their effects in the AGN heating rate.}

\textcolor{black}{Other processes such as turbulent mixing and bulk flows can move energy outside the jet lobes. We expect that these energetics will be captured by our model when radially averaging the AGN heating rate over the cluster. We do this by partitioning the cluster according to the filling factor at each radius; the gas' kinematic profiles (detailed in Sections \ref{Section 3.3.3}-\ref{Section 3.3.4}) will then be given by the volume-weighted mean of the heating-induced and cooling-induced components at each radius. This is equivalent to generating feedback from an effective volumetric energy injection rate of
\begin{equation}\label{effective energy injection}
    \mathcal{Q}_\mathrm{eff}(r) \equiv \sqrt{f(r) \mathcal{H}^2(r) + \left[1 - f(r) \right] C^2(r)},
\end{equation}
which is the root-mean-square profile of the volumetric heating and cooling rates, weighted by their volume contribution. We expect this to capture the energetics transferred kinetically beyond the jet lobes, as the available power is averaged along each radial shell.}

\subsubsection{Parameters of the FR-I jet heating model}\label{Section 3.1.5}

For our FR-I jet heating model, nine input parameters are required to fully describe the heating rate of each jet, on top of the parameters already describing the cluster's environmental and virial parameters (Tables \ref{Table - Environment parameters} and \ref{Table - Virial parameters}, respectively). These nine parameters are summarised in Table \ref{Table - Jet parameters}, along with the physical values we have prescribed for each. We take only the AGN jet duty cycle, $\delta$, as a free parameter hereafter, to investigate how our results depend on AGN jet activity.

\subsection{Evolving the thermodynamic system of the gas}\label{Section 3.2}

\subsubsection{Injecting energy into the gas}\label{Section 3.2.1}

For a thermodynamic system consisting of a homogeneous gas component with thermal pressure, $p_\mathrm{th}$, and volume, $V$, undergoing some energy injection rate, $q$, over time, $\mathrm{d}t$, the gas' state variables will evolve through the differential equation
\begin{equation}\label{Thermodynamic heating}
    q \mathrm{d} t = \left[\frac{1}{\gamma-1}\right]\mathrm{d}p_\mathrm{th}V + \left[\frac{\gamma}{\gamma-1}\right]p_\mathrm{th} \mathrm{d}V,
\end{equation}
which is simply the first law of thermodynamics for an ideal gas. For a non-relativistic ideal gas, its adiabatic index is $\gamma = 5/3$. In terms of the energy injection rate per unit volume, $\mathcal{Q}$, this differential equation becomes
\begin{equation}\label{Heating - Cooling}
    \mathcal{Q} \equiv \frac{q}{V} = \left[\frac{1}{\gamma-1}\right]\frac{\mathrm{d}p_\mathrm{th}}{\mathrm{d}t} + \left[\frac{\gamma}{\gamma-1}\right]\frac{p_\mathrm{th}}{V}\frac{\mathrm{d}V}{\mathrm{d}t},
\end{equation}
where $\mathcal{Q}$ can specify the volumetric energy injection rate due to either the jet's heating or the cluster's cooling. 

For simplicity, we will consider the evolution of a gas component contained in spherical shells with thickness $\ell$, and volume
\begin{equation}\label{Spherical shell}
    V = 4\pi r^2 \ell.
\end{equation}
Taking this shell of volume $V$ and its differential volume $\mathrm{d}V$ into Equation \eqref{Heating - Cooling}, the expression becomes
\begin{equation}\label{Heating - Cooling of spherical shells}
    \mathcal{Q} = \left[\frac{1}{\gamma-1}\right] \frac{\mathrm{d}r}{\mathrm{d}t}  \left[ \frac{\mathrm{d}p_\mathrm{th}}{\mathrm{d}r} + 2 \gamma \frac{p_\mathrm{th}}{r}\right],
\end{equation}
where we have applied the chain rule.

\subsubsection{The propensity for gas motion}\label{Section 3.2.2}

The $\frac{\mathrm{d}r}{\mathrm{d}t}$ term in Equation \eqref{Heating - Cooling of spherical shells} quantifies the propensity\footnote{More specifically, this propensity is proportional to the square-root of the local increase in the gas' kinetic energy, i.e. $\frac{\mathrm{d}r}{\mathrm{d}t} \propto \sqrt{E_\mathrm{kin}}$.} for gas shells to propagate or contract as a result of the energy injection rate at a particular radius. This arises due to the pressure imbalance at that particular radius, subject to the energy injection rate profile, which will require a movement of gas to return the system to hydrostatic equilibrium. We take this term to quantify the gas motion (which can be bulk or turbulent) at a particular radius of the cluster, attributed to the kinetic energy increase of the gas. This solution, which we will refer to as the gas `velocity kick', is
\begin{equation}\label{gas velocity kick}
    v_\mathrm{gas} \equiv \frac{\mathrm{d}r}{\mathrm{d}t} = \frac{(\gamma - 1) \mathcal{Q}}{\left[\frac{\mathrm{d}p_\mathrm{th}}{\mathrm{d}r} + 2\gamma \frac{p_\mathrm{th}}{r}\right]},
\end{equation}
which in our spherical model will depend on radial profiles for the thermal pressure and the volumetric energy injection rate.

To solve Equation \eqref{gas velocity kick}, we must make an important assumption: that the gas remains in approximately a pristine equilibrium state as the system evolves, i.e. assuming that $p_\mathrm{th} \simeq p_\mathrm{eq}$. This approximation will hold if we assume that the movement of gas between spherical shells induces a NTP that is small compared to the thermal pressure of the system; in other words, we assume $p_\mathrm{nt} \ll p_\mathrm{th}$. We confirm the validity of this assumption in Section \ref{Section 6.2.2}. With this assumption, we estimate the gas velocity kick profile as
\begin{equation}\label{simplified gas velocity kick}
    v_\mathrm{gas}(r) \simeq \frac{(\gamma - 1) \mathcal{Q}(r) }{\left[\frac{\mathrm{d}p_\mathrm{eq}(r)}{\mathrm{d}r} + 2\gamma\frac{p_\mathrm{eq}(r)}{r}\right]},
\end{equation}
where we will use the scale-free pressure profile, Equation \eqref{ideal physical baryonic halo pressure}, scaled to the virial parameters in Table \ref{Table - Virial parameters}, and parameterised by the cluster's environmental parameters in Table \ref{Table - Environment parameters}.

\subsubsection{The kinetic energy coupling of the jets to the gas}\label{Section 3.2.3}

In the expressions above, there is no explicit assumption about the amount or type of energy that is coupled to the gas. Instead, this is determined implicitly as the gas system responds to the injection of energy. In particular, when some energy, $q \mathrm{d}t$, is introduced into the system, the propensity for gas to move and gain kinetic energy, i.e. the velocity kick in Equations \eqref{gas velocity kick} and \eqref{simplified gas velocity kick}, encodes a dependence on the background pressure and its radial gradient. This implies that the conversion of injected energy to kinetic energy in the gas component, i.e. the kinetic energy coupling, at any radius of the cluster will depend on the relative size of the denominator of these equations, i.e. $\frac{\mathrm{d}p_\mathrm{eq}(r)}{\mathrm{d}r} + 2\gamma \frac{p_\mathrm{eq}(r)}{r}$, to $(\gamma-1)$ times the volumetric energy injection rate at that radius.

For a given jet outburst of some particular jet power and heating rate, if this denominator is a decreasing function of halo radius, the kinetic energy coupling will be an increasing function of halo radius. In a typical cluster environment, this is always the case: as the gas pressure falls with distance as a power-law, $\frac{\mathrm{d}p_\mathrm{eq}(r)}{\mathrm{d}r} + 2\gamma \frac{p_\mathrm{eq}(r)}{r}$ is a monotonically decreasing function. Therefore, all outbursts in this cluster will have a higher kinetic energy coupling when the jet reaches larger cluster radii. If an outburst is limited in radial extent and ends closer to the core, where this denominator will instead be comparatively larger, the injected energy must instead be converted to thermal energy, as the kinetic energy coupling will be small. This implies that in our model, smaller jet outbursts will predominantly heat the gas, whereas larger jet outbursts are more likely to give a kinetic energy kick to the gas. This implicit energy coupling of the jet outbursts to the gas means that for a particular outburst, the kinetic energy coupling will depend on i) the jet's heating rate, ii) the radius of the cluster in which the jet reaches, and iii) the particular cluster environment at that radius. 

\subsubsection{Separating the AGN heating and the cluster cooling regimes}\label{Section 3.2.4}

Given this model for imparting kinetic energy to the gas, we now consider how energy is injected into separate regions of the cluster when specifying the energy injection rate profile, $\mathcal{Q}(r)$, in Equation \eqref{simplified gas velocity kick}. The heating-induced gas velocity kick, when $\mathcal{Q}\equiv \mathcal{H}$, is
\begin{equation}\label{gas velocity kick due to heating}
    v_{\mathrm{gas, }\mathcal{H}}(r) \simeq  \frac{(\gamma - 1)}{2\pi}\frac{\sqrt{\delta}Q_\mathrm{jet}}{\left[\frac{\mathrm{d}p_\mathrm{eq}(r)}{\mathrm{d}r} + 2\gamma\frac{p_\mathrm{eq}(r)}{r}\right] \int _0 ^{R_\mathrm{outburst}} r^2 f (r) \mathrm{d} r},
\end{equation}
which we assume describes the gas motion inside the jet lobes. The cooling-induced gas velocity kick, when $\mathcal{Q}\equiv C$, is
\begin{equation}\label{gas velocity kick due to cooling}
    v_{\mathrm{gas, }C}(r) \simeq \frac{16(\gamma-1)}{81}\frac{ \Lambda(T_\mathrm{eq}(r))}{\left[\frac{\mathrm{d}p_\mathrm{eq}(r)}{\mathrm{d}r} + 2\gamma\frac{p_\mathrm{eq}(r)}{r}\right]} \left[\frac{\rho_\mathrm{gas}(r)}{\mu m_\mathrm{p}}\right]^2,
\end{equation}
which we assume captures the gas motion outside the jet lobes. In this latter expression, we have again made use of the pristine equilibrium assumption in determining the cooling rate, by approximating this as $C \simeq C_\mathrm{eq}$, as per Equation \eqref{cooling rate - pristine equilibrium}.

\subsection{Describing the non-thermal pressure profile}\label{Section 3.3}

\subsubsection{The non-thermal pressure fraction}\label{Section 3.3.1}

We define the NTP fraction of the intracluster gas as
\begin{equation}
    \mathcal{F} \equiv \frac{p_\mathrm{nt}}{p} = \frac{p_\mathrm{nt}}{p_\mathrm{th} + p_\mathrm{nt}},
\end{equation}
where the total gas pressure is assumed to be the sum of the thermal and NTP components. As pressure is directly proportional to energy density, the NTP fraction is equivalent to a ratio of the corresponding energy densities. 

In our thermodynamic model, NTP will arise as kinetic pressure from induced gas motion. As such, the NTP fraction at any radius within the cluster is simply a ratio of the kinetic to total energy densities of the gas at that radius; this is
\begin{equation}\label{NTP energy densities}
    \mathcal{F} = \frac{u_\mathrm{kin}}{u_\mathrm{th} + u_\mathrm{kin}},
\end{equation}
assuming the total energy density is the sum of thermal, $u_\mathrm{th}$, and kinetic, $u_\mathrm{kin}$, components. 

\subsubsection{The thermal energy density}\label{Section 3.3.2}

For an ideal gas, its thermal energy density  is
\begin{equation}\label{thermal energy density}
    u_\mathrm{th} = \frac{3}{2} \frac{\rho_\mathrm{gas}}{\mu m_\mathrm{p}}k_\mathrm{B} T,
\end{equation}
in terms of the density and temperature of the gas. To determine this energy density in the cluster, we again assume pristine equilibrium, such that it has the radial profile
\begin{equation}\label{thermal energy density - heating n cooling}
    u_\mathrm{th}(r) \simeq \frac{3}{2} \frac{\rho_\mathrm{gas}(r)}{\mu m_\mathrm{p}} k_\mathrm{B} T_\mathrm{eq}(r),
\end{equation}
determined by the temperature profile in Equation \eqref{ideal physical baryonic halo temperature}.

\subsubsection{The kinetic energy density}\label{Section 3.3.3}

For a gas component undergoing non-random motion, its kinetic energy density is
\begin{equation}\label{kinetic energy density}
    u_\mathrm{kin} = \frac{1}{2} \rho_\mathrm{gas} v_{\mathrm{gas}}^2,
\end{equation}
in terms of the gas density and the square of its velocity, $v_\mathrm{gas}$. \textcolor{black}{If this gas motion is described by Equation \eqref{simplified gas velocity kick}, the NTP will be proportional to the square of the volumetric energy injection rate, as $u_\mathrm{kin} \propto v_\mathrm{gas}^2 \propto \mathcal{Q}^2$. For the NTP induced by AGN heating, this implies a proportionality to the square of the AGN's volumetric heating rate, i.e. $p_\mathrm{nt} \propto \mathcal{H}^2$. }

We partition the kinetic energy density into heating and cooling-induced components, which we expect will occur in different regions of the cluster, i.e. inside the jet lobes and in the remainder of the cluster, respectively. We assume that the overall kinematics of the gas can be described by the volume-weighted radial average of the kinetic energy densities; this gives
\begin{equation}\label{kinetic energy density - heating n cooling}
    u_\mathrm{kin}(r) = \frac{1}{2} \rho_\mathrm{gas}(r) \left\{f(r)  v_{\mathrm{gas, }\mathcal{H}}^2(r) + \left[1-f(r)\right] v_{\mathrm{gas, }C}^2(r) \right\},
\end{equation}
where we have separated the heating and cooling regions by the jet's filling factor, $f(r)$, at each radius.

\subsubsection{The NTP fraction as a ratio of square gas velocities}\label{Section 3.3.4}

From the energy densities outlined above, the radial profile for the NTP fraction inside the galaxy cluster is
\begin{equation}\label{NTP for model}
    \mathcal{F}(r) \simeq \frac{1}{\frac{3k_\mathrm{B}T_\mathrm{eq}(r)}{\mu m_\mathrm{p}} \left\{f(r) v_{\mathrm{gas, }\mathcal{H}}^2(r) + \left[1-f(r)\right] v_{\mathrm{gas, }C}^2(r)\right\}^{-1} + 1},
\end{equation}
which depends on the properties of each individual outburst. This fraction is equivalent to a ratio of square gas velocities,
\begin{equation}\label{NTP velocity ratios}
    \mathcal{F}(r) \simeq \frac{v^2_\mathrm{kick}(r)}{v^2_\mathrm{th}(r) + v^2_\mathrm{kick}(r)},
\end{equation}
where 
\begin{equation}\label{effective velocity kick}
    v_\mathrm{kick}(r) \equiv \sqrt{f(r) v_{\mathrm{gas, }\mathcal{H}}^2(r) + \left[1-f(r)\right] v_{\mathrm{gas, }C}^2(r)}
\end{equation}
is the root-mean-square of the heating and cooling-induced velocity kicks, weighted by their volume contribution; and where
\begin{equation}
    v_\mathrm{th} (r) = \sqrt{\frac{3k_\mathrm{B}T_\mathrm{eq}(r)}{\mu m_\mathrm{p}}}
\end{equation}
is the thermal velocity of the gas, when assuming pristine equilibrium.

From these expressions, we can determine the mean radial profile for the NTP fraction, $\left<\mathcal{F}(r)\right>$, when averaging our results over the entire population of jet outbursts in a particular cluster environment.

\begin{figure*}
    \centering
    \includegraphics[width=\textwidth]{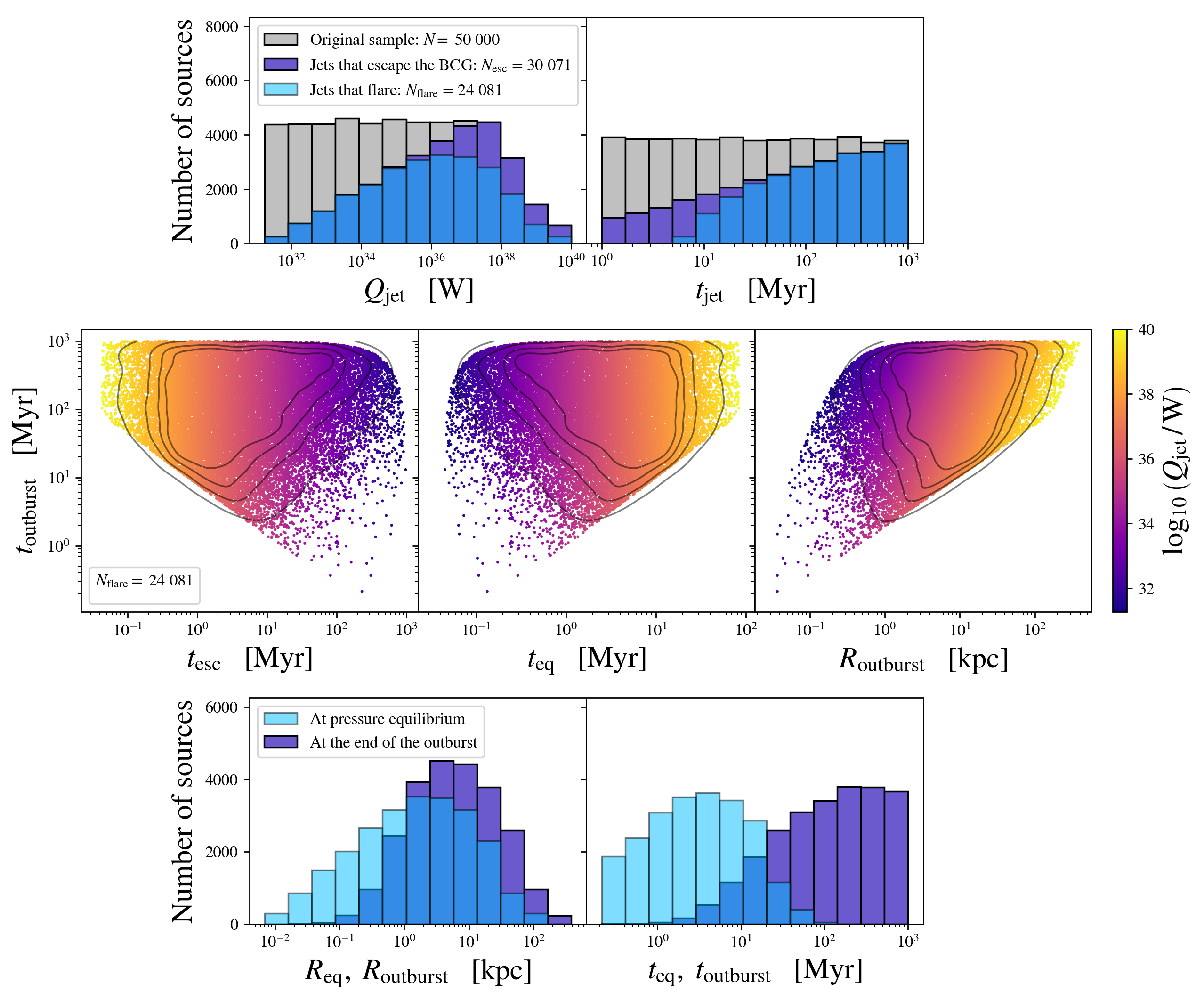}
    \caption{Physical properties of a sample of $N=$ 50 000 jet outbursts evolved in our FR-I model. All input parameters are those given in Tables \ref{Table - Environment parameters}, \ref{Table - Virial parameters} and \ref{Table - Jet parameters}, in a typical cluster environment with $\alpha=1$, $\eta=0.8$ and $\varepsilon=0.5$. 
    \textit{Top row:} histograms showing the sampled jet powers, $Q_\mathrm{jet}$, and jet ages, $t_\mathrm{jet}$, before and after applying the conditions that i) each jet escapes the BCG, reducing the sample to $N_\mathrm{esc} =$ 30 071 jets, and ii) each of these jets reaches pressure equilibrium, reducing the sample to $N_\mathrm{flare} =$ 24 081 jets. \textit{Middle row:} the outburst times, $t_\mathrm{outburst}$, for these jets, plotted against i) their escape times, $t_\mathrm{esc}$, ii) their equilibrium times, $t_\mathrm{eq}$, and iii) their final outburst lengths, $R_\mathrm{outburst}$; these jets are all coloured by their power. The contour lines indicate regions that contain $50\%$, $65\%$, $80\%$ and $95\%$ of the population. \textit{Bottom row:} histograms showing the jet lengths and timescales at pressure equilibrium, $R_\mathrm{eq}, \, t_\mathrm{eq}$, and at the end of the outburst, $R_\mathrm{outburst}, \, t_\mathrm{outburst}$.}
    \label{Fig - histograms panel}
\end{figure*}

\begin{figure*}
    \centering
    \includegraphics[width=\textwidth]{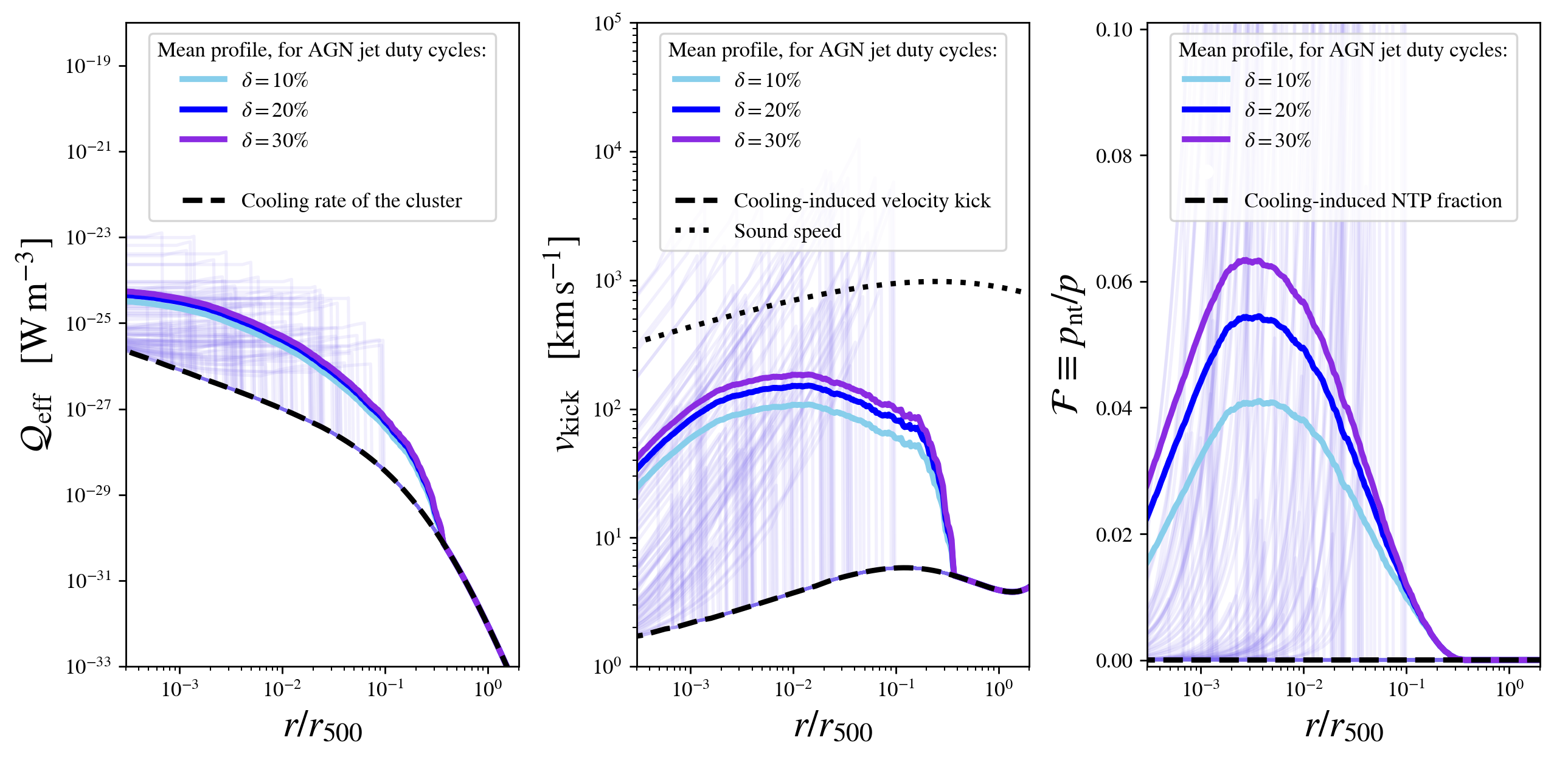}
    \caption{Feedback properties generated by the $N_\mathrm{flare} =$ 24 081 jet outbursts evolved in Figure \ref{Fig - histograms panel}, all as a function of the cluster's scaled halocentric radius, $r/r_{500}$. 
    In each panel, the solid lines are the mean profiles of $N_\mathrm{agn} = N_\mathrm{flare}+N_\mathrm{compact}$ sources, incorporating compact sources with no AGN heating. The colour of these solid lines indicates different choices for the AGN jet duty cycle, $\delta$. The thin faded lines in each panel are the individual trajectories for 100 randomly sampled jet outbursts with $\delta=10\%$. The black dashed lines in each panel show the profiles generated when there is no AGN heating.
    \textit{Left panel:} the cluster's effective volumetric energy injection rates, $\mathcal{Q}_\mathrm{eff}$ (Equation \eqref{effective energy injection}).
    \textit{Middle panel:} the cluster's effective gas velocity kicks, $v_\mathrm{kick}$ (Equation \eqref{effective velocity kick}), compared to the cluster's sound speed, shown by the black dotted line. 
    \textit{Right panel:} the cluster's NTP fractions, $\mathcal{F} \equiv p_\mathrm{nt}/p$, induced by the jet outbursts.
    }
    \label{Fig - heating and cooling panel}
\end{figure*}

\section{THE NON-THERMAL PRESSURE PROFILES}\label{Section 4}

In this section, we present the results of our model, including the mean profiles of the NTP fraction. In Section \ref{Section 4.1}, we provide these profiles for a typical cluster environment. In Section \ref{Section 4.2}, we generalise to a range of cluster environments, including cool cores and non-cool cores, and incorporate large-scale NTP to form complete profiles of the NTP fraction within the cluster. We compare these predictions to those given in the literature.

\subsection{Predictions of our model in a typical cluster}\label{Section 4.1}

\subsubsection{General physical properties of the sample}\label{Section 4.1.1}

Here, we present and discuss some of the key physical properties of our FR-I jet model, when evolved in a typical cluster (given by $\alpha=1$, $\eta=0.8$, $\varepsilon=0.5$; see Section \ref{Section 2.1.2}), with all other parameters given by the discrete values in Tables \ref{Table - Environment parameters} for the remaining environmental parameters, Table \ref{Table - Virial parameters} for the virial parameters, and Table \ref{Table - Jet parameters} for the jet heating parameters. The AGN jet duty cycle, $\delta$, is not required for these predictions. For these parameters, we run $N$ = 50 000 Monte Carlo simulations sampling the probability distributions of jet power and age, creating our source population that we evolve analytically. The key physical properties of this sample are shown in Figure \ref{Fig - histograms panel}.

In the top row of Figure \ref{Fig - histograms panel}, the histograms show our jet population, in terms of their jet power and age, before and after applying the constraints that i) each jet escapes the BCG, and ii) each jet reaches pressure equilibrium. In this environment, we find that $\simeq 60\%$ of these sources escape the BCG, and of these, $\simeq 80\%$ go on to reach pressure equilibrium and flare, forming the population that we consider as jet outbursts, $N_\mathrm{flare}$. The middle row shows various scatter plots for these $N_\mathrm{flare}$ sources, illustrating the relationship between their outburst time and evolutionary properties. As expected, higher-powered sources have shorter escape times and take longer to reach pressure equilibrium, and both higher-powered and longer-lived jets produce more extended outbursts. In the bottom row, the histograms show each jet's final outburst length and outburst time. We see that the final lengths of these outbursts peak around $\simeq$~3-10~kpc. The largest jet in our sample reaches 381~kpc. We overlay this histogram with that showing the jet's equilibrium length and equilibrium time; we see that most of these jets will grow substantially in size by the time the outburst ends. 

\subsubsection{Feedback properties and the mean NTP fraction}\label{Section 4.1.2}

For all of these $N_\mathrm{flare}$ sources that produce outbursts in our cluster, we use the equations detailed in Section \ref{Section 3} to calculate the feedback properties for each jet. In doing so, we must assume a value for the AGN jet duty cycle. We consider separately the values of $\delta=10\%$, $20\%$ and $30\%$, forming three otherwise identical simulated jet populations of size $N_\mathrm{flare}$, but each of varying AGN jet activity. 

Figure \ref{Fig - heating and cooling panel} illustrates these key feedback properties, for each AGN jet duty cycle considered, when evolved in a typical cluster. The mean profiles, shown by the solid colour lines, are calculated over all $N_\mathrm{agn} = N_\mathrm{flare} + N_\mathrm{compact}$ sources, to include the compact source population which have the AGN heating switched off. The thin faded lines trace the trajectories of 100 randomly sampled outbursts when $\delta=10\%$. In the left panel, we show the effective volumetric energy injection rates, $\mathcal{Q}_\mathrm{eff}$, calculated from Equation \eqref{effective energy injection}. Near the cluster's centre, $\mathcal{Q}_\mathrm{eff}$ is dominated by the jet heating rates; in the outskirts, $\mathcal{Q}_\mathrm{eff}$ converges to the cluster's background cooling rate, which is shown by the black dashed line. This background cooling rate is eventually traced by each of the randomly sampled outbursts, as the radial extent of each jet does not reach some cluster radius, resulting in the energy injection rate falling to the cooling rate. In the middle panel, we trace the gas' effective radial velocity kick, $v_\mathrm{kick}$, as in Equation \eqref{effective velocity kick}. In this cluster environment, the mean profiles of these effective velocity kicks never reach the sound speed; this ensures that the mean effective Mach values would be subsonic for all AGN jet duty cycles considered. At large radii, these mean profiles approach the cooling-induced gas velocity kick, Equation \eqref{gas velocity kick due to cooling}, shown by the black dashed line.

Finally, in the right panel of Figure \ref{Fig - heating and cooling panel}, we show the NTP fractions, $\mathcal{F} \equiv p_\mathrm{nt}/p$, for these outbursts. The profile of each individual outburst behaves as an exponentially increasing function of halocentric radius, in some cases reaching values of up to $\mathcal{F}=100\%$, but always rapidly falling to $\mathcal{F}=0\%$ after the outburst; this results in the largest amount of kinetic energy being deposited at the edge of the jet lobes. This is expected from the kinetic energy coupling (Section \ref{Section 3.2.3}). The mean profiles of the sample (again, calculated over all $N_\mathrm{agn}$ sources) have $\left<\mathcal{F}\right>\lesssim 6.3\%$ for the three AGN jet duty cycles shown; this justifies the assumption that the cluster remains well-described, on long timescales, by its pristine equilibrium state, such that $p_\mathrm{nt} \ll p_\mathrm{th}$. We discuss this assumption further in Section \ref{Section 6.2.2}. The black dashed line in this panel, which is zero everywhere, shows the NTP fraction induced by compact sources undergoing only cooling, i.e. those subject to only a cooling-induced velocity kick. As compact sources produce zero NTP in this environment, their inclusion in the mean NTP fraction profile is equivalent to scaling down the mean profile of only extended sources (i.e. the $N_\mathrm{flare}$ population of jets) by a scale factor of $1 - f_\mathrm{compact} = 0.6$.

For the AGN jet duty cycles shown, we find that the cluster's peak mean NTP fractions are $\left<\mathcal{F}\right>\simeq 4.1\%$ when $\delta=10\%$, $\left<\mathcal{F}\right>\simeq 5.4\%$ when $\delta=20\%$, and $\left<\mathcal{F}\right>\simeq 6.3\%$ when $\delta=30\%$. These values are consistent with the NTP fraction measured by \citet{Hitomi2018} in the core of the Perseus cluster. These peaks all occur at $r\simeq$~3~kpc, which \textcolor{black}{at such a small radius, are subject to the influence of the BCG.}\footnote{Whilst we have calibrated the jet's energy budget to escape a particular BCG of radius 2.5~kpc (Sections \ref{Section 2.4.1}-\ref{Section 2.4.3}), as explored in Section \ref{Section 6.1.2} below, changing this radius does not substantially impact our results, and so the BCG's presence was not explicitly modelled in the cluster environment.} \textcolor{black}{Depending on its particular morphology and extent into the cluster environment along the jet axis, the BCG may obscure these peak values.} \textcolor{black}{Despite this possible obscuration, significantly non-zero mean NTP fractions persist into the cluster beyond the 10s of kpc.} These profiles only fall below $\left<\mathcal{F}\right>\lesssim 0.1\%$ at $r\simeq$~275~kpc, indicating the expected extent of the AGN's influence into the cluster medium.

\begin{figure*}
    \centering
    \includegraphics[width=\textwidth]{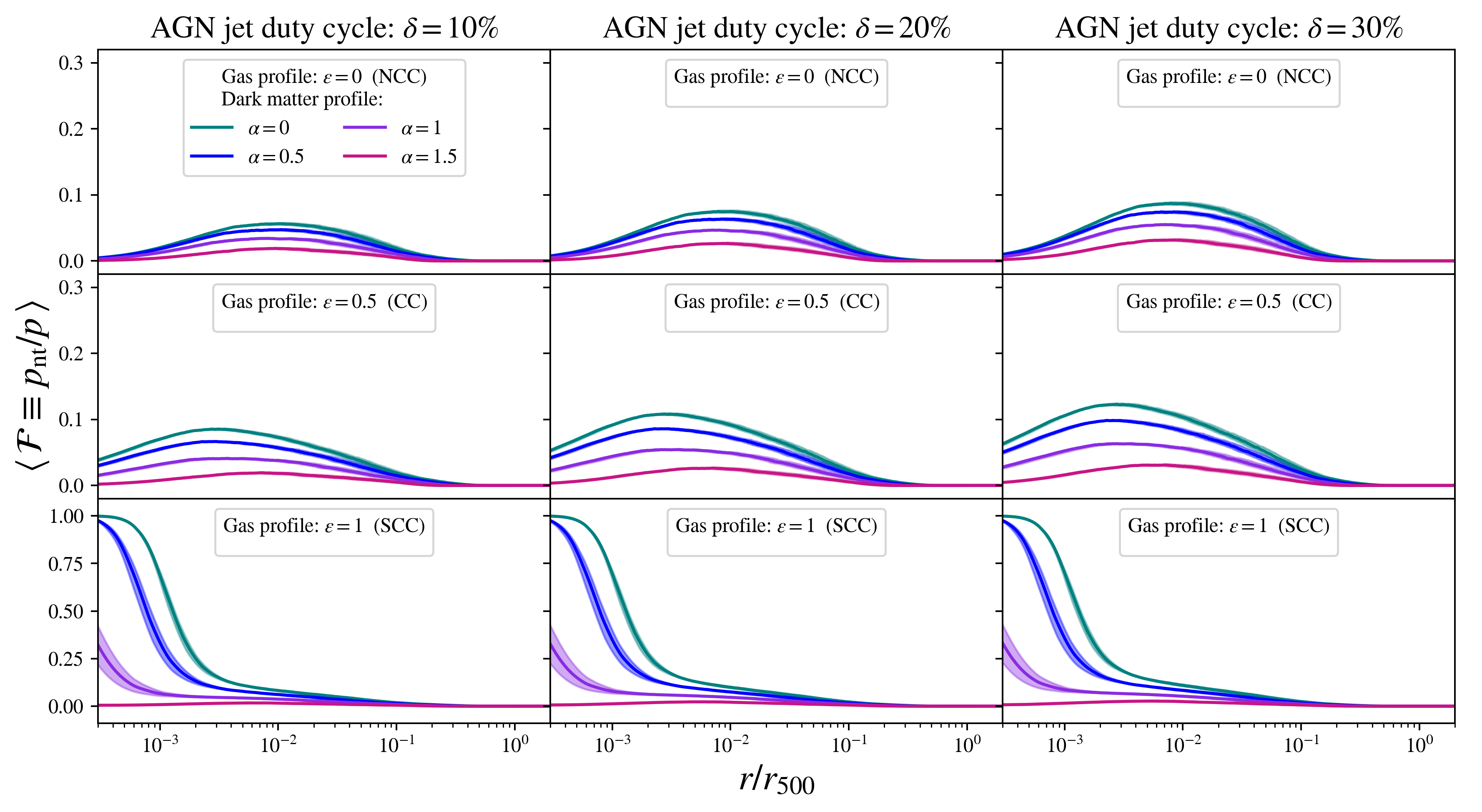}
    \caption{The mean NTP fractions, $\left<\mathcal{F} \equiv p_\mathrm{nt} / p \right>$, traced over the scaled halocentric radius, $r/r_{500}$, predicted by our jet heating model. These profiles are the mean profiles of $N_\mathrm{agn} = N_\mathrm{flare} + N_\mathrm{compact}$ sources, incorporating compact sources with no AGN heating. The AGN jet duty cycle, $\delta$, is varied in each column, and the inner slope of the gas density profile, $\varepsilon$, is varied in each row, for values characteristic of non-cool core (NCC), cool core (CC) and strong cool core (SCC) clusters. Within each box, each colour varies the inner slope of the dark matter density profile, $\alpha$; the solid coloured lines trace the cluster with a fraction of cosmological baryon content of $\eta =0.8$, and the shaded coloured region enclosing each solid line (not visible for all curves) varies this value between $\eta=0.6$ and $\eta=1$. All other parameters in the model take the values given in Tables \ref{Table - Environment parameters}, \ref{Table - Virial parameters} and \ref{Table - Jet parameters}.}
    \label{Fig - NTP profiles in}
\end{figure*}

\begin{figure*}
    \centering
    \includegraphics[width=\textwidth]{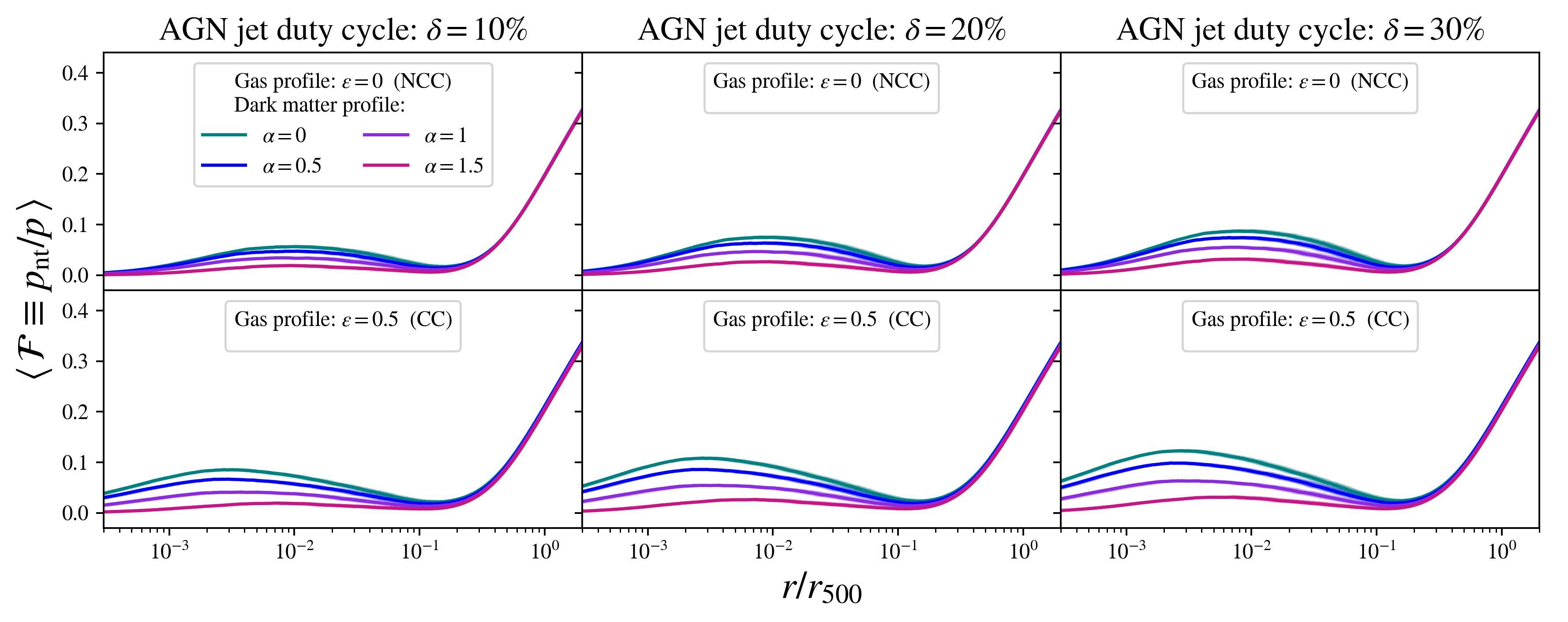}
    \caption{The mean NTP fractions, $\left<\mathcal{F} \equiv p_\mathrm{nt} / p \right>$, traced over the scaled halocentric radius, $r/r_{500}$, when combining the mean NTP fraction profiles from Figure \ref{Fig - NTP profiles in} with the large-scale NTP fraction profiles from \citetalias{Sullivan2024b}. All parameters are the same as in Figure \ref{Fig - NTP profiles in}; the range of values for the inner slope of the gas density profile, $\varepsilon$, is revised to ensure consistency with our modelling assumptions.}
    \label{Fig - NTP profiles in+out}
\end{figure*}

\begin{figure*}
    \centering
    \includegraphics[width=\textwidth]{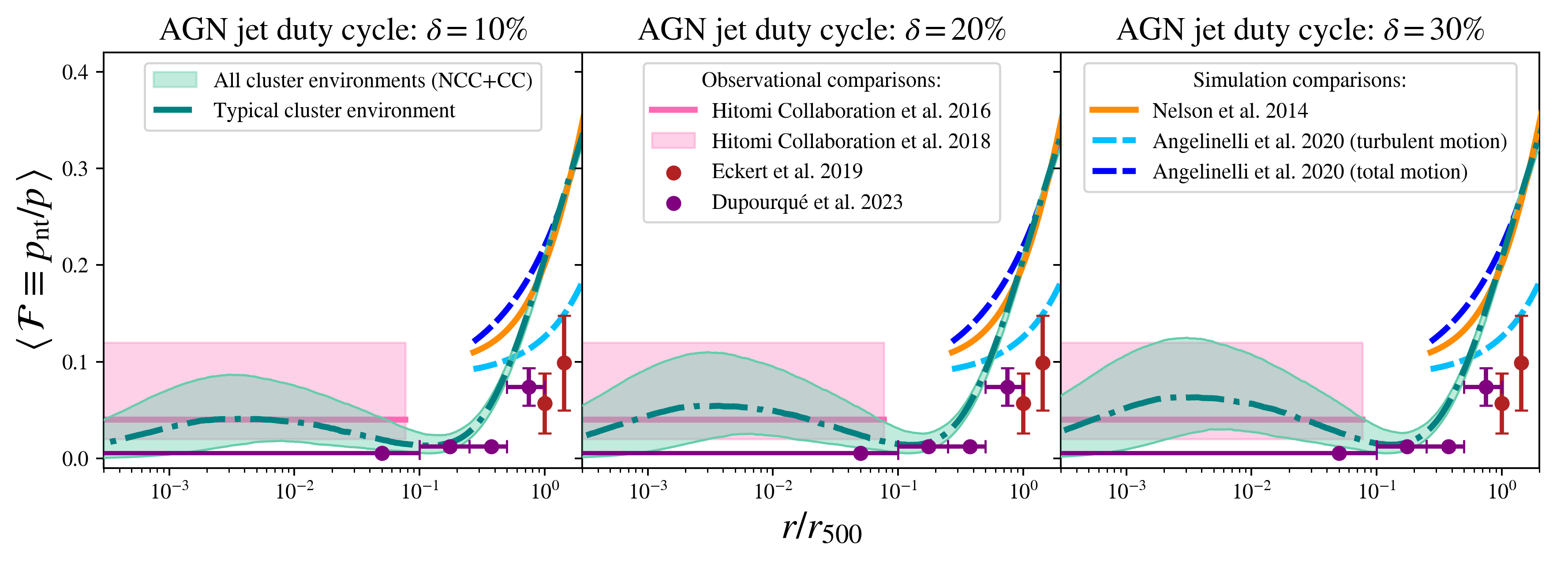}
    \caption{The mean NTP fractions, $\left<\mathcal{F} \equiv p_\mathrm{nt} / p\right>$, traced over the scaled halocentric radius, $r/r_{500}$, predicted in this work for a range of cluster environments (Table \ref{Table - Environment parameters}), including non-cool cores (NCCs) and cool cores (CCs), and for varying AGN jet activity, with the AGN jet duty cycle, $\delta$, varied in each box. The teal shaded region in each box gives the bounds in the mean NTP fraction over all cluster environments; the teal dash-dotted line is the profile for a typical cluster environment. We make comparisons to observational constraints from the \citet{Hitomi2018}, \citet{Eckert2019}, and \citet{Dupourque2023}; as well as to the mean profiles predicted in simulations, from \citet{Nelson2014} and \citet{Angelinelli2020}.}
    \label{Fig - jet NTP predictions}
\end{figure*}

\subsection{Predictions of our model in different clusters}\label{Section 4.2}

\subsubsection{The mean NTP fraction in cool core and non-cool core clusters}\label{Section 4.2.1}

We now repeat this above modelling for different environmental profiles that describe the galaxy cluster, taking into account the range of parameters in Table \ref{Table - Environment parameters}. The resulting mean NTP fractions over these different environments are shown in Figure \ref{Fig - NTP profiles in}, and discussed below.

In the bottom row of Figure \ref{Fig - NTP profiles in}, for cuspy gas profiles with $\varepsilon=1$, we see that the mean NTP fraction profiles grow rapidly toward the cluster's core for all but the $\alpha=1.5$ dark matter halos. When this behaviour occurs, the mean NTP fraction reaches unity, which is higher than the upper limit of $\left<\mathcal{F}\right>=60\%$ that would be expected if the compact sources produced zero NTP (as in a typical cluster, Section \ref{Section 4.1.2}). These $\varepsilon=1$ environments are the only clusters we model that produce non-zero NTP when the AGN heating is switched off. In these environments, regardless of the energy input, the thermal pressure of the gas is so low (see \citetalias{Sullivan2024b}, which discusses the exceedingly cool cores of these clusters) that the kinetic energy coupling becomes exceedingly large near the core. This produces large NTP fractions that are in tension with our modelling assumptions. We note that, as this behaviour occurs inside the region where the BCG would be, this should not be taken as a prediction of our model; this region will be masked or obscured. However, we will consider only gas profiles with inner slopes of $\varepsilon \in [0, 0.5]$ hereafter, to avoid this potential issue of uncontrolled feedback in our results. We leave the treatment of cuspy gas structures for future investigation.

The top two rows of Figure \ref{Fig - NTP profiles in} correspond to `non-cool core' (NCC) clusters, when the gas profile has a core, i.e. $\varepsilon=0$, and `cool core' (CC) clusters, when the gas has a weak cusp, i.e. $\varepsilon=0.5$.\footnote{At this cluster mass, these two gas profiles attain a central entropy of greater than or less than $K(r<0.01r_{500}) =30$~keV cm$^2$ when $\varepsilon=0$ or $\varepsilon=0.5$, respectively, consistent with a NCC versus CC distinction, as in e.g. \citet{Ghirardini2019}; see \citetalias{Sullivan2024b} for the entropy profiles of these clusters.} For these clusters, varying the amount of gas in each cluster produces a negligible change in the mean NTP fractions, as indicated by the barely noticeable shaded intervals enclosing each solid line, which vary the gas content between $\eta=0.6$ and $\eta=1$. In contrast, the shape of the underlying gas density profile, and equivalently the distinction between NCC and CC clusters,\footnote{Although the difference between `strong cool cores' (SCC), i.e. $\varepsilon=1$, and either NCC or CC clusters is clearly more significant in Figure \ref{Fig - NTP profiles in}, we do not take the SCC results to be accurate due to their violation of our modelling assumptions.} is important for predicting NTP, as evident in the increase in the mean NTP fraction between these rows: for any AGN jet duty cycle or dark matter profile, cuspier gas profiles (i.e. cooler cores) attain a higher peak in the mean NTP fraction.

For all NCC and CC clusters in the top two rows of Figure \ref{Fig - NTP profiles in}, the mean NTP fraction exhibits a characteristic peak. This occurs at $r\simeq$~7-8~kpc for NCCs, and $r\simeq$~2-3~kpc (up to $r\simeq$~7~kpc when $\alpha=1.5$) for CCs. \textcolor{black}{The peak values that occur at only $\sim$ a few kpc may be obscured by a BCG, but in all cases, the mean NTP fractions persist at significantly non-zero values beyond several 10s of kpc, beyond potential obscuration.} In a fixed environment, there is a clear relationship between the peak of this mean NTP fraction and the AGN jet duty cycle: increasing $\delta$ increases the peak in $\left<\mathcal{F}\right>$. For NCCs, these peak values are $\left<\mathcal{F}\right> \simeq 5.6\%$ when $\delta=10\%$, $\left<\mathcal{F}\right>\simeq 7.5\%$ when $\delta=20\%$, and $\left<\mathcal{F}\right>\simeq 8.7\%$ when $\delta=30\%$. For CCs, these peak values are $\left<\mathcal{F}\right> \simeq 8.6\%$ when $\delta=10\%$, $\left<\mathcal{F}\right>\simeq 10.8\%$ when $\delta=20\%$, and $\left<\mathcal{F}\right>\simeq 12.3\%$ when $\delta=30\%$. These CC peak values are the upper bounds over all the cluster environments considered in Table \ref{Table - Environment parameters}, after disregarding the $\varepsilon=1$ clusters. These upper bounds occur within the cored, $\alpha=0$, dark matter halos. \textcolor{black}{In contrast, the cuspiest, $\alpha=1.5$, halos are the lower bounds of our results in all gas environments. This is explained by their very high mass densities and large gravitational sinks, forcing the gas to be hotter.} Over all of these NCC and CC environments, the relatively small NTP fractions appear to validate our assumption that $p_\mathrm{nt} \ll p_\mathrm{th}$ in the cluster's core.

\subsubsection{Combining our predictions with the NTP profiles in \citetalias{Sullivan2024b}}\label{Section 4.2.2}

\citetalias{Sullivan2024b} examined the entropy profiles of idealised galaxy clusters modelled over the same parameter space of environments considered in this work (Table \ref{Table - Environment parameters}). By constraining the logarithmic entropy slope, $k(r) \equiv \frac{\mathrm{d}\ln K(r)}{\mathrm{d}\ln r}$, of each environment to meet the value of $k\simeq 1.1$ in the cluster's outskirts, as suggested in both non-radiative simulations \citep[e.g.][]{Voit2005} and X-ray observations \citep[e.g.][]{Ghirardini2019} of clusters, the required NTP fraction was solved as a function of cluster radius. This analytical model from \citetalias{Sullivan2024b} was not predictive of the NTP inside the cluster's core, which was controlled via a normalisation of the overall profile; instead, these profiles provide larger-scale predictions nearer the cluster's outskirts. The NTP profile suggested by \citetalias{Sullivan2024b}, after being normalised to zero NTP in the core, provided strong agreement with the mean NTP profiles predicted in numerical simulations \citep[from][]{Nelson2014, Angelinelli2020}; we take these profiles as the ideal template in which to add in the predictions of this work. With zero NTP in the cluster's core, the question of AGN feedback is left entirely to our new model.

In Figure \ref{Fig - NTP profiles in+out}, we radially add the mean NTP fraction profiles from Figure \ref{Fig - NTP profiles in} to the corresponding NTP profiles from \citetalias{Sullivan2024b} (which are scale-free in the cluster regime, and not calibrated to a particular halo mass), for each of the same prescriptions of environmental parameters of NCCs and CCs. These combined profiles show that a transition occurs between the dominant non-thermal contribution to each cluster's hydrostatic equilibrium state, whereby the NTP predicted by our `inside-out' model approaches zero at approximately the same point that the NTP from the `outside-in' \citetalias{Sullivan2024b} model becomes non-zero. This manifests as a `dip' in the NTP profile that occurs around $r \simeq 100-200$~kpc, which is visible for most cluster environments in Figure \ref{Fig - NTP profiles in+out}, but most evident in clusters experiencing higher AGN jet activity. 

\subsubsection{Comparing the mean NTP fractions to observations}\label{Section 4.2.3}

We now compare our predicted mean NTP fraction profiles, when generalising our results over all NCC and CC cluster environments,  to constraints provided in the literature. By taking the maximum range permitted by these profiles, in Figure \ref{Fig - NTP profiles in+out}, at each radius and for each choice of the AGN jet duty cycle, we form the interval bounds in the mean NTP fraction shown in Figure \ref{Fig - jet NTP predictions}. We indicate the mean profile for a typical cluster by the teal dash-dotted line in each panel. We compare our results to the mean NTP fraction profiles determined from gas turbulence in hydrodynamic simulations, \citet{Nelson2014} and \citet{Angelinelli2020}, with the latter study employing two separate definitions of gas turbulence: based on the total velocity dispersion (`total motion'), or the smoothed local velocity field (`turbulent motion'). As discussed in \citetalias{Sullivan2024b}, our results show excellent agreement with these numerical predictions \citep[specifically the total motion definition for][]{Angelinelli2020}. \textcolor{black}{The observational predictions in this region, from \citet{Eckert2019} and \citet{Dupourque2023}, remain in tension with our NTP profile and these numerical predictions.}

Our results in the cluster core strongly suggest that the AGN feedback imparted by energetic jet outbursts does not induce a large amount of NTP to the surrounding gas, which is in good agreement with the inference provided by observations. Constraints given by \citet{Dupourque2023} in the cluster core find $\mathcal{F} \lesssim 1.5\%$ out to a cluster radius of $r\simeq 0.5r_{500}$, which is consistent with our mean profiles within $r\simeq 50-200$~kpc. The \citet{Hitomi2018} observations of the Perseus cluster, shown by the pink intervals, provide a strong comparison to our results for all AGN jet duty cycles considered. The \citet{Hitomi2016} value of $\mathcal{F}\simeq 4\%$ appears to favour AGN jet duty cycles of the order $\delta=10\%$, particularly in the environment of a typical cluster. 

\textcolor{black}{The first results from XRISM (published whilst this work has been going through submission) have provided a strong case that the NTP in cores of galaxy clusters is small but non-zero. These observations include $\mathcal{F}\simeq 2\%$ in Abell 2029, a relaxed AGN cluster \citep{XRISM_Abell2029}; $\mathcal{F}\simeq 3\%$ in Centaurus, an active AGN cluster \citep{XRISM_Centaurus}; $\mathcal{F}\simeq 4.5\%$ in Hydra-A, which hosts powerful FR-I jets \citep{XRISM_HydraA}; and $\mathcal{F}\simeq 3\%$ in Coma, a merging cluster without significant AGN jet activity \citep{XRISM_Coma}. These results suggest that AGN activity, in the form of jet outbursts, has a potentially limited, but non-zero, influence on the gas turbulence in the cluster core and that other processes (e.g. mergers) may also contribute to observed NTP fractions. We highlight that our model confirms that jet feedback induces only limited amounts of NTP into the cluster core, with the NTP fraction increasing by only $\sim$ a few $\%$ when the AGN jet activity increases. This may be at the level of variance that these first XRISM results suggest.}

\section{The Connection to AGN Jet Activity}\label{Section 5}

We now investigate the relationship between the mean NTP fraction and the AGN jet duty cycle, which provides an important practical application of our model. We quantify this relationship in Section \ref{Section 5.1}, and then apply this to the Perseus cluster in Section \ref{Section 5.2}. This allows us to predict Perseus' AGN jet duty cycle, which we compare with independent observational constraints on its recent AGN jet activity. 

\subsection{Non-thermal pressure and the AGN jet duty cycle}\label{Section 5.1}

\subsubsection{Fitting a relationship to our results}\label{Section 5.1.1}

We can now quantify how the NTP predicted in our model is related to the AGN jet duty cycle. As is evident in Figures \ref{Fig - NTP profiles in}, \ref{Fig - NTP profiles in+out} and \ref{Fig - jet NTP predictions}, the height of the peak mean NTP fraction increases when increasing the AGN jet duty cycle, $\delta$, for all NCC and CC clusters. When fixing all other parameters in our model, we assume that this peak mean NTP fraction is described by some power series expansion in $\delta$. We validate this assumption by the fact that the NTP fraction generated by any jet outburst, at any location in the cluster subject to its heating, is approximately
\begin{equation}\label{NTP with duty cycle}
    \mathcal{F}(r) \simeq \frac{f(r)v^2_{\mathrm{gas,}\mathcal{H}}(r)}{v^2_\mathrm{th}(r) + f(r)v^2_{\mathrm{gas,}\mathcal{H}}(r)} = \frac{\delta}{\mathcal{A}(r) + \delta},
\end{equation}
for some positive-valued radial function, $\mathcal{A}(r)$. This approximation arises by simplifying Equation \eqref{NTP velocity ratios} when $f(r) v^2_{\mathrm{gas,}\mathcal{H}} \gg \left[1-f(r)\right]v^2_{\mathrm{gas,}C}$, i.e. where heating substantially dominates over cooling in the cluster, and by taking $v_{\mathrm{gas, }\mathcal{H}} \propto \sqrt{\delta}$ as per Equation \eqref{gas velocity kick due to heating}. We use this approximation, Equation \eqref{NTP with duty cycle}, to posit that the mean NTP fraction of the sample, at locations in the cluster where AGN heating dominates, will also encode a dependence on $\delta$. As such, we choose to fit the peak mean NTP fraction to a power series expansion of leading-order $\delta$, such that
\begin{equation}\label{power series fit}
    \left<\mathcal{F}\right>_\mathrm{peak} \simeq k_0 + k_1 \sqrt{\delta} + k_2 \delta
\end{equation}
for some set of coefficients, $k_0, k_1, k_2 \in \mathbb{R}$, that will encompass the properties of the cluster environment and the mean jet feedback properties at the location of this peak. We fix the constant term to be $k_0 = 0$, as we expect $\left<\mathcal{F}\right> \simeq 0$ when there is no heating, i.e. when $\delta=0\%$, by the expectation that cooling cannot produce non-zero NTP in the environments we are modelling (see Section \ref{Section 4.2.1}).

\begin{figure}
    \centering
    \includegraphics[width=0.5\textwidth]{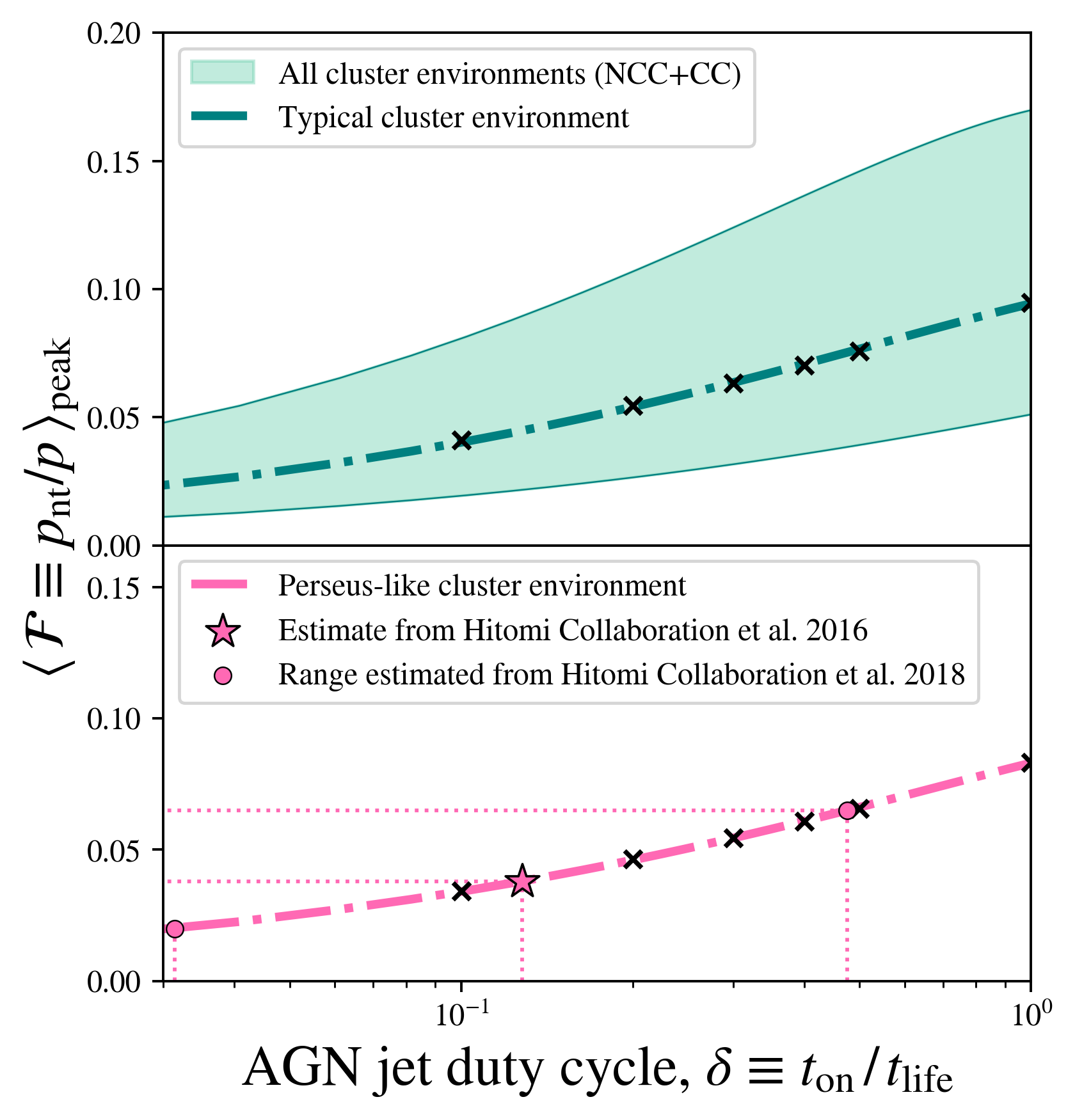}
    \caption{The relationship between the peak mean NTP fraction and the AGN jet duty cycle that arises in our jet heating model. In each panel, the black crosses are the points fit to the power series expansion (Table \ref{Table - Power series fits}). \textit{Top panel:} the predicted relationship over a range of cluster environments (Table \ref{Table - Environment parameters}), with the teal dash-dotted line the relationship we find for a typical cluster environment. \textit{Bottom panel:} the predicted relationship for a Perseus-like cluster, which we use to estimate Perseus' AGN jet duty cycle from its NTP observations. The pink star gives the AGN jet duty cycle we estimate using the \citet{Hitomi2016} NTP measurement; the pink circles span the range allowed by the \citet{Hitomi2018} range in NTP values.}
    \label{Fig - peak NTP v duty cycle}
\end{figure}

We determine the remaining coefficients for each of the cluster environments in Figure \ref{Fig - NTP profiles in+out} by fitting the peak values of $\left<\mathcal{F}\right>$ for different values of $\delta$ to Equation \eqref{power series fit}. The interval prescribed by these fits is shown in the top panel of Figure \ref{Fig - peak NTP v duty cycle}, with the relationship that a typical cluster falls on shown by the teal dash-dotted line; this fit takes the values of $k_1 = 0.142$ and $k_2= -0.048$. Along this relationship, the points we have fitted are shown in black crosses, clearly validating our choice of this particular power series expansion. The coefficients for all other environments we consider in this work are detailed in Table \ref{Table - Power series fits}. In each case, the $\sqrt{\delta}$ term dominates the fit. Of note, we do not vary the baryon content, $\eta$, in these fits, as the peak values of $\left<\mathcal{F}\right>$ are not sensitive to this parameter, as evident in Figures \ref{Fig - NTP profiles in} and \ref{Fig - NTP profiles in+out}. 

This relationship illustrates one of the key findings of our work: if the mean NTP fraction is observationally estimated in the core of a galaxy cluster undergoing AGN jet activity, then, under the assumption that this NTP is primarily due to its jet outbursts, its AGN jet duty cycle can be constrained. Furthermore, if the gas profiles in the cluster's core are known, then in theory its AGN jet duty cycle could be directly measured. This relationship would be applicable to individual clusters with frequent and known past AGN activity, or to a large sample of clusters of similar mass and environment so that the estimated mean NTP fraction would be representative of a large number of jet outbursts.

\renewcommand{\arraystretch}{1.5}
\begin{table}
\begin{tabular}{ |p{3.2cm}||p{1.4cm}|p{1.4cm}| }
\hline
\rowcolor{lightgray!30}
 Cluster environment: & \multicolumn{1}{|c|}{$\boldsymbol{k_1}$} & \multicolumn{1}{|c|}{$\boldsymbol{k_2}$} \\
\hline\hline
\rowcolor{white} \cellcolor{lightgray!30} $\boldsymbol{\alpha=0, \, \varepsilon=0} \,$ (NCC) & 0.196 & -0.068 \\ 
\hline
\rowcolor{white} \cellcolor{lightgray!30} $\boldsymbol{\alpha=0.5, \, \varepsilon=0} \,$  (NCC) & 0.165 & -0.053 \\ 
\hline
\rowcolor{white} \cellcolor{lightgray!30} $\boldsymbol{\alpha=1, \, \varepsilon=0} \,$  (NCC) & 0.120 & -0.035 \\ 
\hline
\rowcolor{white} \cellcolor{lightgray!30} $\boldsymbol{\alpha=1.5, \, \varepsilon=0} \,$  (NCC) & 0.066 & -0.015 \\ 
\hline
\rowcolor{white} \cellcolor{lightgray!30} $\boldsymbol{\alpha=0, \, \varepsilon=0.5} \,$  (CC) & 0.295 & -0.125 \\ 
\hline
\rowcolor{white} \cellcolor{lightgray!30} $\boldsymbol{\alpha=0.5, \, \varepsilon=0.5} \,$ (CC) & 0.230 & -0.090 \\ 
\hline
\rowcolor{white} \cellcolor{lightgray!30} $\boldsymbol{\alpha=1, \, \varepsilon=0.5} \,$  (CC) \newline \footnotesize{\text{(`Typical cluster')}} & 0.142 & -0.048  \\ 
\hline
\rowcolor{white} \cellcolor{lightgray!30} $\boldsymbol{\alpha=1.5, \, \varepsilon=0.5} \,$  (CC)& 0.067 & -0.018 \\ 
\hline
\rowcolor{white} \cellcolor{lightgray!30} {Perseus-like cluster} \newline \footnotesize{\text{\citep{Churazov2003}}} & 0.119 & -0.036 \\ 
\hline
\end{tabular}
\caption{The coefficients, $k_1$ and $k_2$, fit to the power series in Equation \eqref{power series fit}. These fits are given for different cluster environments, including non-cool cores (NCCs) and cool cores (CCs) (Table \ref{Table - Environment parameters}), as well as for a Perseus-like cluster environment (Section \ref{Section 5.2.2}).}
\label{Table - Power series fits}
\end{table}

\subsection{The AGN jet duty cycle of the Perseus cluster}\label{Section 5.2}

\subsubsection{Observational constraints on Perseus' AGN jet duty cycle}\label{Section 5.2.1}

As of writing, the \textit{Hitomi} measurement of the Perseus cluster's central NTP fraction provides the logical real galaxy cluster in which we can apply this relationship between the peak NTP and the AGN jet duty cycle. The Perseus cluster is known to have experienced frequent outbursts in its recent history \citep[e.g.][]{Fabian2000, Fabian2006}, such that its central NTP is expected to be representative of several jet outbursts that have occurred during the AGN's lifecycle. Further, X-ray emission from Perseus has been used to fit its intracluster gas profiles \citep[e.g.][]{Churazov2003}, such that its environment is well-constrained.

Observational signatures of remnant bubbles in Perseus' core can be used to estimate its recent AGN jet activity. \citet{Fabian2003} found the distance between ripples in the core of Perseus to be $\lambda \simeq 11$~kpc, with a sound speed at the bubble surface of $c_\mathrm{s} \simeq 1.17\times 10^6$~m~s$^{-1}$. Assuming each ripple is representative of a separate outburst, the time between successive outbursts, $\lambda/c_\mathrm{s}$, is then $t_\mathrm{on}+t_\mathrm{off}\simeq 9.2$~Myr. The time it takes for a bubble (or, equivalently, a jet lobe) to inflate, i.e. the outburst time, can be estimated by applying a simple jet lobe expansion model to an observed bubble. We use the relationship given in \citet{KaiserAlexander1997}:
\begin{equation}\label{bubble radius growth}
    R(t) = c_1 \left[\frac{Q_\mathrm{jet}}{\rho_\mathrm{gas}}\right]^{1/5}t^{3/5},
\end{equation}
where $c_1$ is a dimensionless constant of order unity. Taking the ratio of the lobe length, $R(t)$, and the lobe velocity, $\dot{R}(t)$, at the end of the outburst, we expect the outburst time to be
\begin{equation}
    t_\mathrm{outburst} = \frac{3}{5} \frac{R_\mathrm{bubble}}{\dot{R}_\mathrm{bubble}}.
\end{equation}
\citet{ShabalaAlexander2009} state a bubble radius of $R_\mathrm{bubble}\simeq$~6~kpc in Perseus; \citet{Graham2008} give a Mach value of $\mathcal{M} \simeq 1.2$ at the bubble's shock front, corresponding to a bubble velocity of $\dot{R}_\mathrm{bubble} \simeq 1.2 c_\mathrm{s}$. From these values, we estimate $t_\mathrm{outburst}\simeq 2.5$~Myr. This suggests Perseus has an AGN jet duty cycle of $\delta\simeq 27\%$. 

For this recent jet outburst, taking the number density of gas to be $n_\mathrm{gas} \simeq 0.116$~cm$^{-3}$ at the edge of the bubble \citep[as in][]{ShabalaAlexander2009}, using Equation \eqref{bubble radius growth} we estimate that its jet power would be $Q_\mathrm{jet}\simeq 7\times 10^{36}$~W when $c_1 \simeq 1.5$, which is consistent with the constraints in jet power given in \citet{ShabalaAlexander2009}. This consistency suggests that $\delta \simeq 27\%$ is a sensible estimate for Perseus' AGN jet duty cycle.

\subsubsection{Our prediction for Perseus' AGN jet duty cycle}\label{Section 5.2.2}

To estimate Perseus' AGN jet duty cycle using the model presented here, we simulate a new cluster environment that matches observational fits for the Perseus cluster. We normalise this cluster to a virial radius of $r_{500} = 1.3$~Mpc \citep{Urban2014}, implying a virial mass of $M_{500} =~5.8~\times ~10^{14}~\mathrm{M}_\odot$. We use X-ray fits obtained for the electron number density,
\begin{equation}\label{Perseus gas density}
    \frac{n_\mathrm{e}}{\mathrm{m^{-3}}} =  \frac{3.9\times 10^4}{\left[1 + \left(\frac{r}{80~\mathrm{kpc}}\right)^2\right]^{1.8}} + \frac{4.05\times 10^3}{\left[1 + \left(\frac{r}{280~\mathrm{kpc}}\right)^2\right]^{0.87}},
\end{equation}
and temperature,
\begin{equation}\label{Perseus temperature}
    \frac{T}{\mathrm{keV}} = \frac{7\left[1 + \left(\frac{r}{100~\mathrm{kpc}}\right)^3\right]}{\left[2.3 + \left(\frac{r}{100~\mathrm{kpc}}\right)^3\right]},
\end{equation}
as given in \citet{Churazov2003}. From the above profiles, we can recover the thermal pressure of the gas, which we continue to assume is the hydrostatic equilibrium solution (i.e. we assume pristine equilibrium), balancing the cluster's underlying gravitational potential, which we do not need to prescribe.

\begin{figure}
    \centering
    \includegraphics[width=0.36\textwidth]{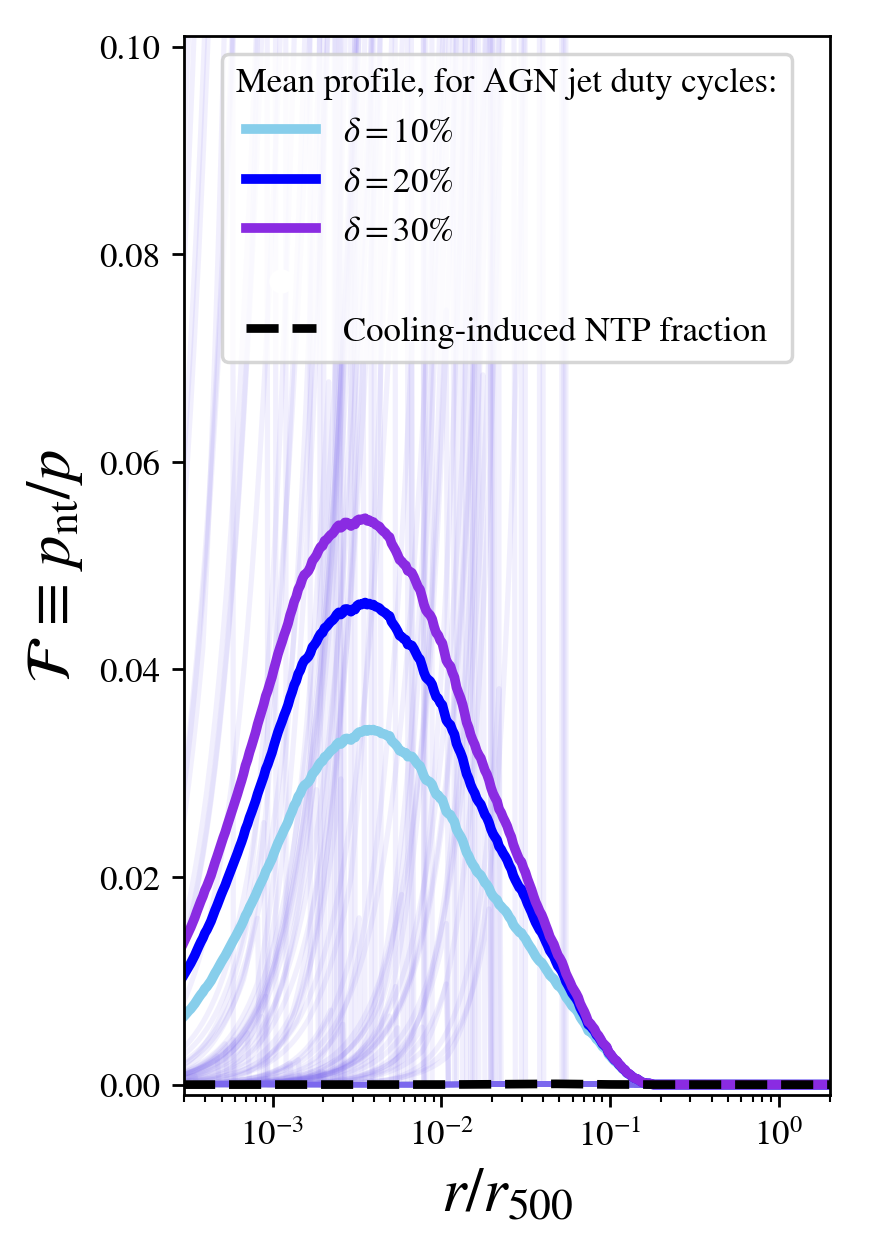}
    \caption{The NTP fractions, $\mathcal{F} \equiv p_\mathrm{nt}/p$, traced over the halocentric radius, $r/r_{500}$, generated in a Perseus-like cluster, simulated by the gas profiles given in \citet{Churazov2003}. The solid lines are the mean profiles of $N_\mathrm{agn} = N_\mathrm{flare}+N_\mathrm{compact}$ sources, incorporating compact sources with no AGN heating. The colour of these solid lines indicates different choices for the AGN jet duty cycle, $\delta$. The thin faded lines are the individual trajectories for 100 randomly sampled jet outbursts with $\delta=10\%$. The black dashed line shows the profile generated when there is no AGN heating.}
    \label{Fig - Perseus NTP panel}
\end{figure}

We run our jet feedback model as before for $N =$ 50 000 Monte Carlo simulations in this Perseus-like environment, for a range of AGN jet duty cycles. The resulting mean NTP fraction profiles are shown in Figure \ref{Fig - Perseus NTP panel}, which obey the same general form as found in previous cluster environments. Compared to the typical cluster, the Perseus-like cluster attains a smaller mean NTP fraction for a given AGN jet duty cycle. Fitting the predicted peak mean NTP fractions in this environment to Equation \eqref{power series fit}, we find $k_1 = 0.119$ and $k_2 = -0.036$. This is remarkably close to the fit given for the NCC cluster with $\alpha=1$ and $\varepsilon=0$, as per Table \ref{Table - Power series fits}. The relationship for Perseus is shown in the bottom panel of Figure \ref{Fig - peak NTP v duty cycle} in the pink dash-dotted line. 

From this relationship, we use the \textit{Hitomi} measurements to estimate Perseus' AGN jet duty cycle. We apply our relationship to i) the single \citet{Hitomi2016} value of $\mathcal{F} \simeq 4\%$, and ii) the isotropic motion-inferred values in the \citet{Hitomi2018}, that give the range of $2\% \lesssim \mathcal{F} \lesssim 7\%$.\footnote{We cannot consider the upper bound of $\mathcal{F} \lesssim 11-12\%$ given in \cite{Hitomi2018} if the gas is undergoing sloshing motion, as these values are higher than we can model in our simulated Perseus-like cluster.} We also note that \textit{Hitomi} measured the fraction of NTP to thermal pressure, i.e. $x\equiv p_\mathrm{nt}/p_\mathrm{th}$, so that our corresponding NTP fractions will be slightly smaller, by $\mathcal{F} = (1+\frac{1}{x})^{-1}$, which we take into account in this analysis.

Using the single \citet{Hitomi2016} value, we predict Perseus' AGN jet duty cycle to be $\delta \simeq 13\%$. For the \citet{Hitomi2018} range in values, we predict Perseus' AGN jet duty cycle to be within $3\% \lesssim \delta \lesssim 48\%$. This range is in good agreement with observational evidence of Perseus' AGN jet activity. We caution that these constraints are very sensitive to the measured NTP values, due to the $\sqrt{\delta}$ term driving the relationships in Figure \ref{Fig - peak NTP v duty cycle}. The observational capabilities of XRISM will allow higher precision in these NTP measurements. The promising agreement of our model with independent measurements of Perseus' AGN jet activity suggests that our work could apply to these future observations. This will allow estimates to be made for the past AGN activity of other real clusters.

\section{MODEL CAVEATS}\label{Section 6}

We now examine the dependence of our results on key assumptions made during our modelling, which form the main caveats of our work. We discuss the sensitivity of our results to the jet modelling, in terms of the jet power distribution and the escape time calibration, in Section \ref{Section 6.1}. We then quantify the impact of violating the underlying pristine equilibrium assumption, by applying perturbations to the cluster's hydrostatic equilibrium state; this is discussed in Section \ref{Section 6.2}.

\subsection{Dependence on the jet modelling}\label{Section 6.1}

\begin{figure*}
    \centering
    \includegraphics[width=\textwidth]{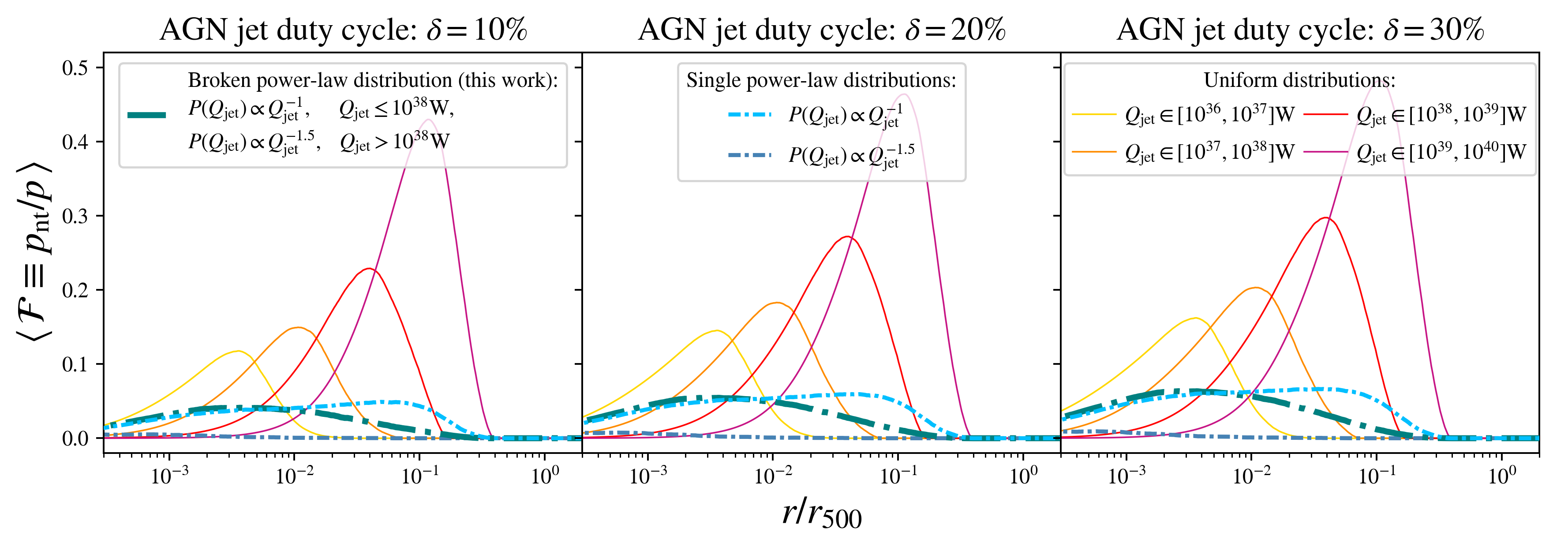}
    \caption{The mean NTP fractions, $\left<\mathcal{F} \equiv p_\mathrm{nt} / p\right> $, traced over the scaled halocentric radius, $r/r_{500}$, in a typical cluster environment, but for different assumptions about the distribution of jet powers, $P(Q_\mathrm{jet})$. The teal dash-dotted line is the result given in our work, using a broken power-law distribution. The light blue and steel blue densely dash-dotted lines give the profiles using single power-law distributions, of slopes $s_Q = 1$ and $s_Q = 1.5$, respectively. The thin coloured lines give the profiles generated from uniform distributions of jet power within magnitude intervals. The AGN jet duty cycle, $\delta$, is varied in each box.}
    \label{Fig - jet power sensitivity}
\end{figure*}

\begin{figure*}
    \centering
    \includegraphics[width=\textwidth]{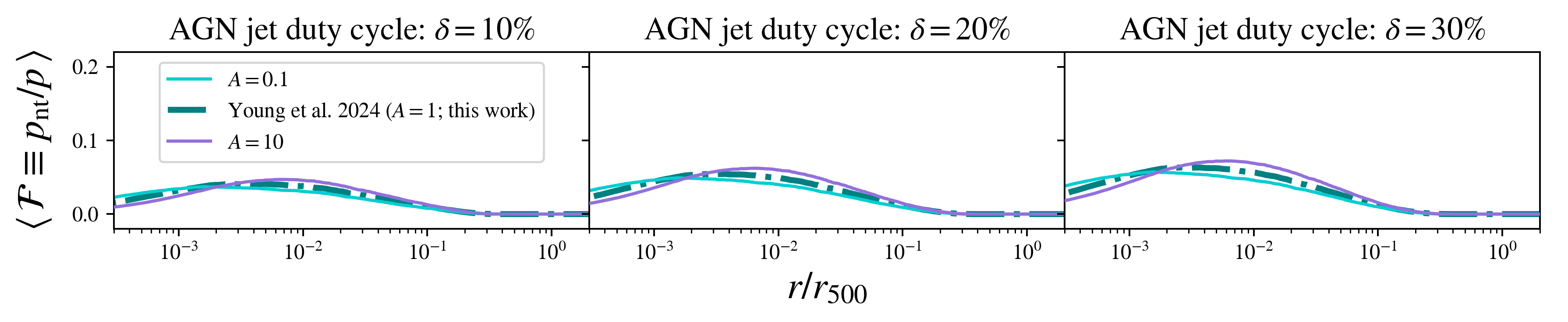}
    \caption{The mean NTP fractions, $\left<\mathcal{F} \equiv p_\mathrm{nt} / p\right> $, traced over the scaled halocentric radius, $r/r_{500}$, in a typical cluster environment, but for different assumptions about the brightest cluster galaxy (BCG) used in the escape time calibration. The teal dash-dotted line is the result given in our work, calibrated to the simulated jet outbursts in \citet{Young2024}. The thin turquoise and indigo lines consider a change by a multiplicative factor of $A$ in the BCG gas mass (or radius) from that in \citet{Young2024}. The AGN jet duty cycle, $\delta$, is varied in each box.}
    \label{Fig - calibration sensitivity}
\end{figure*}

\begin{figure*}
    \centering
    \includegraphics[width=\textwidth]{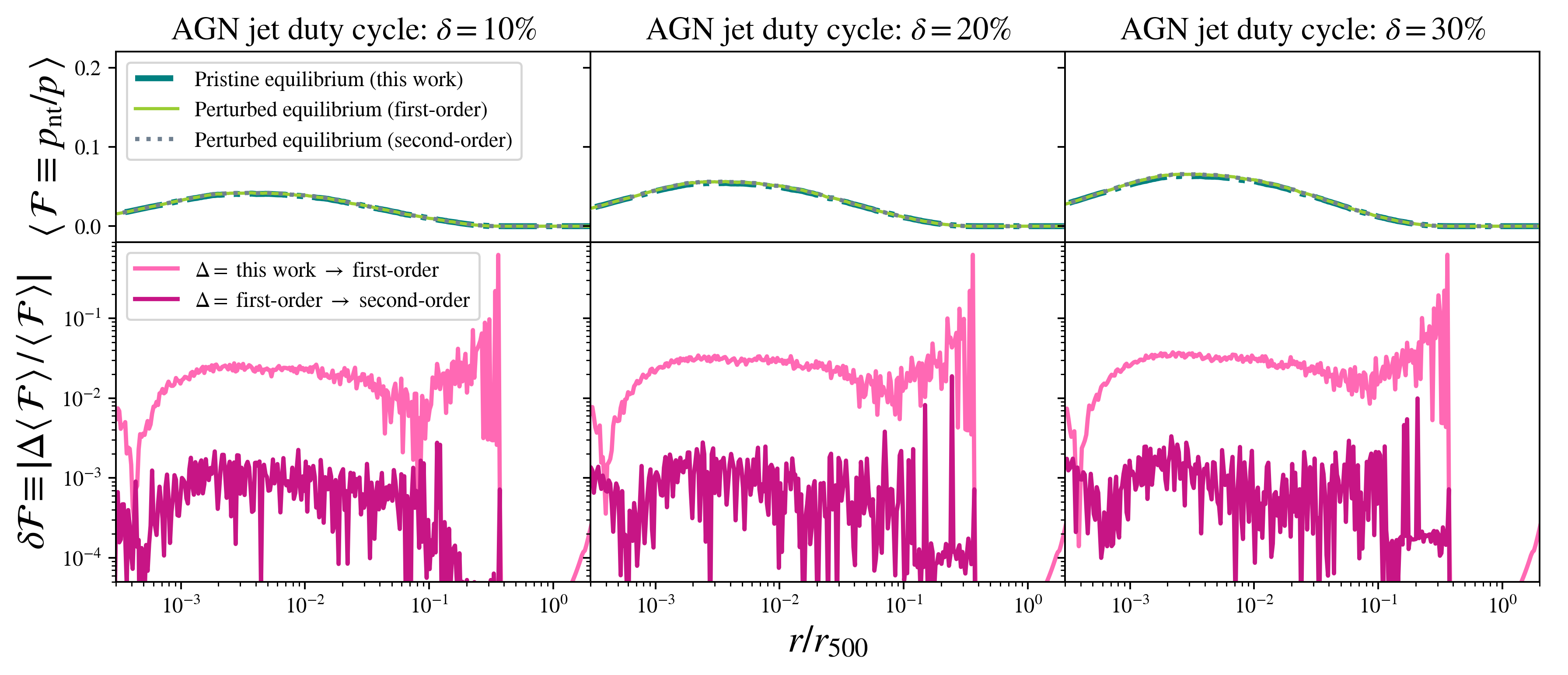}
    \caption{The sensitivity of the predictions of this work to perturbing the hydrostatic equilibrium state of the cluster, in a typical cluster environment. All profiles are traced over the scaled halocentric radius, $r/r_{500}$. The AGN jet duty cycle, $\delta$, is varied in each column. \textit{Top panels:} the mean NTP fractions, $\left<\mathcal{F} \equiv p_\mathrm{nt} / p\right> $, before and after each perturbation. The teal dash-dotted line is the result given in our work, i.e. for pristine equilibrium; the thin yellow-green line perturbs the pressure balance to account for the mean NTP fractions predicted in this work; the grey dotted line perturbs the pressure balance again to account for the mean NTP fractions predicted after the first-order perturbation. \textit{Bottom panels:} the fractional change in the mean NTP fraction, $\delta\mathcal{F} \equiv \left|\Delta \left<\mathcal{F}\right>/\left<\mathcal{F}\right>\right|$, between each of these perturbations. }
    \label{Fig - pristine equilibrium assumption}
\end{figure*}

\subsubsection{Sensitivity to the jet power distribution}\label{Section 6.1.1}

In this study, we have assumed that the AGN's jet outbursts follow a broken power-law distribution in jet powers, as given in Equation \eqref{broken power law jet power distribution}. In this section, we compare our mean NTP fraction profiles to those generated by single power-law distributions in jet power, e.g. of the form in Equation \eqref{jet power distribution}, when in a typical cluster environment. This allows us to examine the differences introduced to our model when following the single power-law slopes suggested by \citet{Shabala2020} and \citet{Quici2024}. We show these comparisons in Figure \ref{Fig - jet power sensitivity}. To illustrate the relative contribution of different jet powers to each profile, we also show the profiles generated by uniform distributions in jet power, sampled linearly within magnitude intervals.

For the broken power-law assumed in this work, we see that the mean NTP fraction is dominated by sources of jet power within $36 \lesssim \log_{10} (Q_\mathrm{jet}/\mathrm{W}) \lesssim 38$. Importantly, we find that higher-powered sources, of $\log_{10} (Q_\mathrm{jet}/\mathrm{W}) \gtrsim 39$, contribute little if at all to these mean profiles, justifying our use of an FR-I model. Higher-powered sources would generally be FR-II in type, which would evolve differently from what we have modelled; if these higher-powered sources did contribute significantly to our results, a combined FR-I and FR-II evolutionary model would likely be needed.

For the single power-law with slope $s_Q = 1$, as suggested by \citet{Shabala2020}, we see that the dominant sources contributing to the mean NTP fraction profiles are those of jet power within $37 \lesssim \log_{10} (Q_\mathrm{jet}/\mathrm{W}) \lesssim 39$. For this distribution, an FR-II model would likely be required to more accurately evolve a moderate fraction of these higher-powered sources. When compared to the results of our broken power-law distribution, we see that the peak of the mean NTP fraction is smoothed, reaching further out in the cluster. This only marginally changes the peak mean NTP fraction, although this peak occurs much further out into the cluster. This larger spread in the mean NTP fraction further into the cluster is in tension with the results from \citet{Dupourque2023}.

For the case of the single power-law with slope $s_Q = 1.5$, as suggested by \citet{Quici2024}, we see that the mean NTP fraction profiles are essentially zero everywhere throughout the cluster, as the population is dominated by lower-powered sources that produce negligible NTP. This indicates that such a steep distribution in jet powers may not be entirely representative of the AGN population, but might hold only for certain jet powers (as in, e.g. our broken power-law distribution). The choice in the distribution of jet powers is of strong importance to our model, highlighting the sensitivity of our results to constraints on the jet power distribution of AGN galaxies.

\subsubsection{Sensitivity to the BCG escape time calibration}\label{Section 6.1.2}

In ensuring that our jet outbursts were generated with physically-motivated outburst times and minimum jet powers, we applied a BCG escape criterion to our jet sample, calibrated to simulations given in \citet{Young2024}. We now consider the effect of changing the BCG's properties on our resulting mean NTP fraction profiles. This occurs via Equations \eqref{calibrated escape time} and \eqref{calibrated jet power}, where we can scale the BCG gas mass, $M_\mathrm{gas}$, by some multiplicative factor of $A$ from the value given in \citet{Young2024}.\footnote{This is mathematically equivalent to an identical change by a factor $A$ in the enclosing BCG radius, $r_\mathrm{gas}$; we interpret these changes in terms of the BCG gas mass for conciseness.} 

In Figure \ref{Fig - calibration sensitivity}, we consider an order of magnitude decrease, i.e. $A=0.1$, and increase, i.e. $A=10$, in the BCG gas mass used to calibrate our model. When increasing (decreasing) the BCG gas mass, the minimum jet power required to escape the BCG is increased (decreased), resulting in the mean profile being dominated by higher (lower)-powered jet sources that reach larger (smaller) sizes. This results in the radial location of the peak mean NTP fraction changing by $\sim$ a few kpc, but with the peak NTP fraction changing by less than a percentage of the total pressure. We conclude that changes to the BCG properties have only minor impacts on our results in this calibration. 

\textcolor{black}{We expect that the BCG will only impact our results by obscuring the peak mean NTP within its radial extent, although this is not explicitly modelled in our work. Incidentally, the shift in these mean NTP fraction profiles, outward for more massive BCGs and inward for less massive BCGs, appears to compensate for this expected obscuration when varying the mass (and equivalently, the radius) of these galaxies.}

\subsection{Dependence on the hydrostatic equilibrium state}\label{Section 6.2}

\subsubsection{The pristine equilibrium assumption}\label{Section 6.2.1}

Throughout this study, we assumed that the cluster is always in a state of pristine equilibrium, explicitly ignoring any NTP contribution to the cluster's pressure balance. This assumption appears three times in our model:

\begin{enumerate}
    \item In the gas velocity kick approximation, Equation \eqref{simplified gas velocity kick}, where we assumed that the gas' thermal pressure and its radial derivative could be approximated as
    $p_\mathrm{th} \simeq p_\mathrm{eq}$ and $ \frac{\mathrm{d}p_\mathrm{th}}{\mathrm{d}r} \simeq \frac{\mathrm{d}p_\mathrm{eq}}{\mathrm{d}r}$, respectively. 
    \item In the cooling-induced gas velocity kick, Equation \eqref{gas velocity kick due to cooling}, where we assumed the cooling function $\Lambda (T) \simeq \Lambda (T_\mathrm{eq})$ in the cooling rate. 
    \item In the NTP fraction, Equations \eqref{NTP for model} and \eqref{NTP velocity ratios}, where we assumed that the thermal energy density and thermal velocity emerged from a gas temperature of $T \simeq T_\mathrm{eq}$. 
\end{enumerate}
All three of these assumptions will be broken by the predictions of our jet model, which introduces NTP into the cluster. Whilst assumptions (i) and (iii) are of high importance, assumption (ii) can be considered unimportant, as the contribution of any cooling-induced gas velocity to the NTP fraction is completely negligible (as shown in Section \ref{Section 4.2.1} for the environments we model). 

\subsubsection{Perturbing the cluster's hydrostatic equilibrium state}\label{Section 6.2.2}

We now investigate how the predictions given in this work hold up when perturbing the cluster's hydrostatic equilibrium state beyond pristine equilibrium, by introducing NTP into the system. Following AGN jet activity and the introduction of NTP, if the system returns to a local hydrostatic equilibrium balance, the thermal pressure of the gas must be reduced to account for the emergent NTP, which we account for by using our mean NTP fraction profiles. We assume this occurs on timescales less than that between outbursts. This will reduce the thermal pressure and temperature of the gas by
\begin{equation}
    p_\mathrm{th}(r) = p_\mathrm{eq}(r) \left[1 - \left<\mathcal{F}(r)\right>\right]
\end{equation}
and
\begin{equation}
    T(r) = T_\mathrm{eq}(r) \left[1 - \left<\mathcal{F}(r)\right>\right],
\end{equation}
respectively, when assuming that the gas density profile remains unchanged. 

We re-run the model for this new pressure balance, keeping the same AGN jet duty cycles, to generate new predictions for the mean NTP fractions. We apply this perturbation twice in a typical cluster environment, as shown in the top row of Figure \ref{Fig - pristine equilibrium assumption}; we also show the resulting fractional change in these profiles, $\delta \mathcal{F} \equiv \left| \Delta \left<\mathcal{F}\right> /\left<\mathcal{F}\right>\right|$, between each perturbation, in the bottom row.

As evident from these panels, the changes in the mean NTP fractions between these perturbations are minor. At the NTP peak, the fractional changes induced by the first-order perturbations are: $\delta \mathcal{F} \simeq 2.5\%$ when $\delta=10\%$, $\delta \mathcal{F} \simeq 3.3\%$ when $\delta=20\%$, and $\delta \mathcal{F} \simeq 3.4\%$ when $\delta=30\%$. Between the first-order and the second-order perturbations, the fractional change is over an order of magnitude smaller, at $\delta \mathcal{F} \simeq 0.1\%$. This suggests that the iterative process of updating the pressure balance and recalculating the mean NTP fraction is convergent, requiring at most a first-order perturbation to be within $\sim$ a few $\%$ accuracy of the expected solution.

This perturbation analysis shows how a cluster may evolve when re-establishing hydrostatic equilibrium after accruing NTP from repeated AGN jet activity, assuming the induced NTP is long-lived and that pressure re-configuration occurs on short timescales. The convergence of these solutions indicates that there may be a steady-state solution to the pressure balance in the cores of active galaxies.

\section{CONCLUSION}\label{Section 7}

In this study, we have presented a statistics-based analytical approach to predict the mean feedback properties in the cores of galaxy clusters subject to AGN jet activity. \textcolor{black}{By employing observational constraints on the power-law probability distributions in jet powers and jet ages, we generated a population of jet outbursts which we embedded inside an analytical cluster environment. We evolved each of these outbursts in time by assuming an FR-I jet evolutionary model, transitioning from a ballistic to a flaring jet phase when the jet reached pressure equilibrium with its environment. When coupling the injected energy to the surrounding gas, by applying the first law of thermodynamics to spherical radial shells, we predicted the radial kinetic energy density increase induced by each outburst. We then averaged these results over the outburst population, when varying both the AGN jet duty cycle and the environmental properties of the cluster.} 

The key findings from this study are summarised below.

\begin{itemize}
    \item Over a range of cluster environments, including non-cool cores and cool cores (Table \ref{Table - Environment parameters}), we found that the mean profile of the non-thermal pressure (NTP) fraction, $\mathcal{F} \equiv p_\mathrm{nt}/p$, induced on the gas due to AGN jet activity, is bounded by theoretical upper limits. These are
        \begin{itemize}
            \item $\left<\mathcal{F}\right> \lesssim 9\%$, when $\delta=10\%$,
            \item $\left<\mathcal{F}\right> \lesssim 11\%$, when $\delta=20\%$,
            \item $\left<\mathcal{F}\right> \lesssim 12\%$, when $\delta=30\%$,
        \end{itemize}
    where $\delta \equiv t_\mathrm{on}/t_\mathrm{life}$ is the AGN jet duty cycle. 

    \item \textcolor{black}{For an observationally-inferred cluster environment, with an NFW dark matter profile, a weakly cusped gas profile and typical baryon content, we identified the peak values of $\left<\mathcal{F}\right> \simeq 4-6\%$ when $\delta =10-30\%$; these peak values occur at a cluster radius of $r\simeq$~3~kpc.}
    
    \item When $\delta$ is constant, we found that $\left<\mathcal{F}\right>$ increases for cuspier gas density profiles (i.e. cooler cores); this suggests that such environments are more susceptible to kinetic AGN feedback. 

    \item We combined our predictions with the larger-scale NTP constraints from earlier work (\citetalias{Sullivan2024b}), predicting a complete radial NTP fraction profile for galaxy clusters. 

    \item Our predictions in the cluster core are in good agreement with observations, such as the \citet{Hitomi2018}, \citet{Dupourque2023}, \textcolor{black}{and the first available XRISM observations,} which all suggest that NTP is not the dominant mode of AGN feedback. 

    \item We predicted a relationship between $\delta$ and the peak value of $\left<\mathcal{F}\right>$ within a given cluster environment. This suggests that the AGN jet activity of real clusters could be inferred from an observed NTP fraction and its intracluster gas profiles.

    \item We applied our model to a Perseus-like simulated cluster, finding $\delta \simeq 13\%$ or the range $3\% \lesssim \delta \lesssim 48\%$ when using the \textit{Hitomi} measurements of NTP in Perseus' core. This range is in strong agreement with independent evidence of Perseus' AGN jet activity. 

    \item We found that the predictions of this work are robust when perturbing the cluster's hydrostatic equilibrium state to account for the predicted NTP contribution. For maintenance-mode AGN feedback, the long-term mean NTP fraction in the cluster's core can be modelled as a steady-state solution of this pressure balance. 
\end{itemize}

We have shown that the NTP produced by energetic jet outbursts in the cores of both non-cool core and cool core galaxy clusters is, on average, always small, implying that kinetic AGN feedback does not significantly alter a cluster's hydrostatic equilibrium balance. Does the relationship we find between NTP and the AGN jet duty cycle hold in simulations of jet outbursts in AGN host galaxies? And if so, what does this imply for upcoming observational surveys, which will measure the NTP in many more galaxy clusters? \textcolor{black}{We plan to investigate this relationship between NTP and AGN jet activity further. We hope that when applied in conjunction with measurements of the current state of the ambient medium and the AGN jet activity, our models will help unveil the past AGN activity of real clusters from their observed NTP fractions. This is currently possible with the science being achieved by XRISM, making our work a relevant and realistic application to these and future observations.}

\vspace{1cm}

\section*{Acknowledgements}

 AS acknowledges the support of the Australian Government Research Training Program Fees Offset; the Bruce and Betty Green Postgraduate Research Scholarship; and The University Club of Western Australia Research Travel Scholarship. AS and CP acknowledge the support of the ARC Centre of Excellence for All Sky Astrophysics in 3 Dimensions (ASTRO 3D), through project number CE170100013. SS and CP acknowledge support from the ARC through project DP240102970. \textcolor{black}{We thank our reviewer Filip Huško for his helpful and constructive comments.} We also thank Nicole Thomas and Martin Bourne for their helpful insights.

\vspace{1cm}

\section*{Data availability}

The Python code for the jet outburst model used in this study is publicly available at: \\
https://github.com/andrew-sullivan-2022/Jet\_NTP\_Model

\vspace{1cm}
 


\bibliographystyle{mnras}
\bibliography{example} 



\bsp	
\label{lastpage}
\end{document}